%
%
\input harvmac %
\input epsf
%
%
%
%
%

%
%
%
%
\newif\ifdraft

\noblackbox
\catcode`\@=11
\newif\iffrontpage
%
\ifx\answ\bigans
\def\titleft{\titsm}
\magnification=1200\baselineskip=15pt plus 2pt minus 1pt
%
\advance\hoffset by-0.075truein
\hsize=6.15truein\vsize=600.truept\hsbody=\hsize\hstitle=\hsize
\else\let\lr=L
\def\titleft{\titla}
\magnification=1000\baselineskip=14pt plus 2pt minus 1pt
%
\vsize=6.5truein
\hstitle=8truein\hsbody=4.75truein
\fullhsize=10truein\hsize=\hsbody
\fi
\parskip=4pt plus 10pt minus 4pt

\font\titla=cmr10 scaled\magstep3
\font\tenmss=cmss10
\font\absmss=cmss10 scaled\magstep1
\newfam\mssfam
\font\footrm=cmr8  \font\footrms=cmr5
\font\footrmss=cmr5   \font\footi=cmmi8
\font\footis=cmmi5   \font\footiss=cmmi5
\font\footsy=cmsy8   \font\footsys=cmsy5
\font\footsyss=cmsy5   \font\footbf=cmbx8
\font\footmss=cmss8
\def\footfont{\def\rm{\fam0\footrm}
\textfont0=\footrm \scriptfont0=\footrms
\scriptscriptfont0=\footrmss
\textfont1=\footi \scriptfont1=\footis
\scriptscriptfont1=\footiss
\textfont2=\footsy \scriptfont2=\footsys
\scriptscriptfont2=\footsyss
\textfont\itfam=\footi \def\it{\fam\itfam\footi}
\textfont\mssfam=\footmss \def\mss{\fam\mssfam\footmss}
\textfont\bffam=\footbf \def\bf{\fam\bffam\footbf} \rm}
\def\tenpoint{\def\rm{\fam0\tenrm}
\textfont0=\tenrm \scriptfont0=\sevenrm
\scriptscriptfont0=\fiverm
\textfont1=\teni  \scriptfont1=\seveni
\scriptscriptfont1=\fivei
\textfont2=\tensy \scriptfont2=\sevensy
\scriptscriptfont2=\fivesy
\textfont\itfam=\tenit \def\it{\fam\itfam\tenit}
\textfont\mssfam=\tenmss \def\mss{\fam\mssfam\tenmss}
\textfont\bffam=\tenbf \def\bf{\fam\bffam\tenbf} \rm}
\ifx\answ\bigans\def\abstractfont{\tenpoint}\else
\def\abstractfont{\def\rm{\fam0\absrm}
\textfont0=\absrm \scriptfont0=\absrms
\scriptscriptfont0=\absrmss
\textfont1=\absi \scriptfont1=\absis
\scriptscriptfont1=\absiss
\textfont2=\abssy \scriptfont2=\abssys
\scriptscriptfont2=\abssyss
\textfont\itfam=\bigit \def\it{\fam\itfam\bigit}
\textfont\mssfam=\absmss \def\mss{\fam\mssfam\absmss}
\textfont\bffam=\absbf \def\bf{\fam\bffam\absbf}\rm}\fi
%
\def\f@@t{\baselineskip10pt\lineskip0pt\lineskiplimit0pt
\bgroup\aftergroup\@foot\let\next}
\setbox\strutbox=\hbox{\vrule height 8.pt depth 3.5pt width\z@}
\def\vfootnote#1{\insert\footins\bgroup
\baselineskip10pt\footfont
\interlinepenalty=\interfootnotelinepenalty
\floatingpenalty=20000
\splittopskip=\ht\strutbox \boxmaxdepth=\dp\strutbox
\leftskip=24pt \rightskip=\z@skip
\parindent=12pt \parfillskip=0pt plus 1fil
\spaceskip=\z@skip \xspaceskip=\z@skip
\Textindent{$#1$}\footstrut\futurelet\next\fo@t}
\def\Textindent#1{\noindent\llap{#1\enspace}\ignorespaces}
\def\footnote#1{\attach{#1}\vfootnote{#1}}%

\def\foot{\attach\footsymbolgen\vfootnote{\footsymbol}}
\let\footsymbol=\star
\newcount\lastf@@t           \lastf@@t=-1
\newcount\footsymbolcount    \footsymbolcount=0
\def\footsymbolgen{\relax\footsym
\global\lastf@@t=\pageno\footsymbol}
\def\footsym{\ifnum\footsymbolcount<0
\global\footsymbolcount=0\fi
{\iffrontpage \else \advance\lastf@@t by 1 \fi
\ifnum\lastf@@t<\pageno \global\footsymbolcount=0
\else \global\advance\footsymbolcount by 1 \fi }
\ifcase\footsymbolcount \fd@f\star\or
\fd@f\dagger\or \fd@f\ast\or
\fd@f\ddagger\or \fd@f\natural\or
\fd@f\diamond\or \fd@f\bullet\or
\fd@f\nabla\else \fd@f\dagger
\global\footsymbolcount=0 \fi }
\def\fd@f#1{\xdef\footsymbol{#1}}
\def\space@ver#1{\let\@sf=\empty \ifmmode #1\else \ifhmode
\edef\@sf{\spacefactor=\the\spacefactor}
\unskip${}#1$\relax\fi\fi}
 \def\attach#1{\space@ver{\strut^{\mkern 2mu #1}}\@sf}
%
\newif\ifnref
\def\rrr#1#2{\relax\ifnref\nref#1{#2}\else\ref#1{#2}\fi}
\def\ldf#1#2{\begingroup\obeylines
\gdef#1{\rrr{#1}{#2}}\endgroup\unskip}
\def\nrf#1{\nreftrue{#1}\nreffalse}

\nreffalse
\def\refout{\listrefs}
%
\def\eqn#1{\xdef #1{(\secsym\the\meqno)}
\writedef{#1\leftbracket#1}%
\global\advance\meqno by1\eqno#1\eqlabeL#1}
\def\eqnalign#1{\xdef #1{(\secsym\the\meqno)}
\writedef{#1\leftbracket#1}%
\global\advance\meqno by1#1\eqlabeL{#1}}
%
\def\chap#1{\newsec{#1}}
\def\chapter#1{\chap{#1}}
\def\sect#1{\subsec{{ #1}}}
\def\section#1{\sect{#1}}
\def\\{\ifnum\lastpenalty=-10000\relax
\else\hfil\penalty-10000\fi\ignorespaces}
\def\note#1{\leavevmode%
\edef\@@marginsf{\spacefactor=\the\spacefactor\relax}%
\ifdraft\strut\vadjust{%
\hbox to0pt{\hskip\hsize%
\ifx\answ\bigans\hskip.1in\else\hskip-.1in\fi%
\vbox to0pt{\vskip-\dp
\strutbox\sevenbf\baselineskip=8pt plus 1pt minus 1pt%
\ifx\answ\bigans\hsize=.7in\else\hsize=.35in\fi%
\tolerance=5000 \hbadness=5000%
\leftskip=0pt \rightskip=0pt \everypar={}%
\raggedright\parskip=0pt \parindent=0pt%
\vskip-\ht\strutbox\noindent\strut#1\par%
\vss}\hss}}\fi\@@marginsf\kern-.01cm}
\def\titlepage{%
\frontpagetrue\nopagenumbers\abstractfont%
\hsize=\hstitle\rightline{\vbox{\baselineskip=10pt%
{\abstractfont\pubnum}}}\pageno=0}
\frontpagefalse
\def\pubnum{}
\def\pdate{\number\month/\number\yearltd}
\def\makefootline{\iffrontpage\vskip .27truein
\line{\the\footline}
\vskip -.1truein\leftline{\vbox{\baselineskip=10pt%
{\abstractfont\pdate}}}
\else\vskip.5cm\line{\hss \tenrm $-$ \folio\ $-$ \hss}\fi}
\def\title#1{\vskip .7truecm\titlestyle{\titleft #1}}
\def\titlestyle#1{\par\begingroup \interlinepenalty=9999
\leftskip=0.02\hsize plus 0.23\hsize minus 0.02\hsize
\rightskip=\leftskip \parfillskip=0pt
\hyphenpenalty=9000 \exhyphenpenalty=9000
\tolerance=9999 \pretolerance=9000
\spaceskip=0.333em \xspaceskip=0.5em
\noindent #1\par\endgroup }
\def\autskip{\ifx\answ\bigans\vskip.5truecm\else\vskip.1cm\fi}
\def\author#1{\vskip .7in \centerline{#1}}

\def\address#1{\ifx\answ\bigans\vskip.2truecm
\else\vskip.1cm\fi{\it \centerline{#1}}}
\def\abstract#1{
\vskip .3in\vfil\centerline
{\bf Abstract}\penalty1000
{{\smallskip\ifx\answ\bigans\leftskip 2pc \rightskip 2pc
\else\leftskip 5pc \rightskip 5pc\fi
\noindent\abstractfont \baselineskip=12pt
{#1} \smallskip}}
\penalty-1000}
\def\endpage{\tenpoint\supereject\global\hsize=\hsbody%
\frontpagefalse\footline={\hss\tenrm\folio\hss}}
\def\ack{\goodbreak\vskip1.5cm\noindent {\bf Acknowledgements}}


\def\bfone{\relax{\rm 1\kern-.35em 1}}
\def\inbar{\vrule height1.5ex width.4pt depth0pt}
\def\IC{\relax\,\hbox{$\inbar\kern-.3em{\mss C}$}}
\def\ID{\relax{\rm I\kern-.18em D}}
\def\IF{\relax{\rm I\kern-.18em F}}
\def\IH{\relax{\rm I\kern-.18em H}}
\def\II{\relax{\rm I\kern-.17em I}}
\def\IN{\relax{\rm I\kern-.18em N}}
\def\IP{\relax{\rm I\kern-.18em P}}
\def\IQ{\relax\,\hbox{$\inbar\kern-.3em{\rm Q}$}}
\def\IR{\relax{\rm I\kern-.18em R}}
\font\cmss=cmss10 \font\cmsss=cmss10 at 7pt
\def\ZZ{\relax\ifmmode\mathchoice
{\hbox{\cmss Z\kern-.4em Z}}{\hbox{\cmss Z\kern-.4em Z}}
{\lower.9pt\hbox{\cmsss Z\kern-.4em Z}}
{\lower1.2pt\hbox{\cmsss Z\kern-.4em Z}}\else{\cmss Z\kern-.4em
Z}\fi}
\def\a{\alpha} \def\b{\beta} 
 \def\c{\gamma}

\def\cN{{\cal N}}

\def\t1{t_1}
\def\t2{t_2}
\def\t3{t_3}
\def\t4{t_4}
\def\t5{t_5}
\def\nup#1({Nucl.\ Phys.\ $\us {B#1}$\ (}
\def\plt#1({Phys.\ Lett.\ $\us  {#1}$\ (}
\def\cmp#1({Comm.\ Math.\ Phys.\ $\us  {#1}$\ (}
\def\prp#1({Phys.\ Rep.\ $\us  {#1}$\ (}
\def\prl#1({Phys.\ Rev.\ Lett.\ $\us  {#1}$\ (}
\def\prv#1({Phys.\ Rev.\ $\us  {#1}$\ (}
\def\mpl#1({Mod.\ Phys.\ \Let.\ $\us  {#1}$\ (}
\def\tit#1|{{\it #1},\ }
%

%

\def\ni{\noindent}
\def\tilde{\widetilde}

\def\us#1{\underline{#1}}

\def\Coe#1.#2.{{#1\over #2}}
\def\coeff#1#2{\relax{\textstyle {#1 \over #2}}\displaystyle}
\def\coe#1.#2.{\relax{\textstyle {#1 \over #2}}\displaystyle}

\def\to{\rightarrow}
\def\notin{\hbox{{$\in$}\kern-.51em\hbox{/}}}

 \def\ie{{i.e.}}
\catcode`\@=12
%
\def\note#1{\ifdraft {\bf [#1]} \else\fi}
\def\cref{\ifdraft {\bf [check refs]}\else\fi}

\def\undertext#1{\vtop{\hbox{#1}\kern 1pt \hrule}}
\def\subsection#1{\vskip 0.5cm\undertext{\rm #1}}
\def\Ftwo{F_{II}}
\def\Fhet{F_{H}}

\def\nh{N_H}
\def\nv{N_V}
\def\nt{N_T}

\def\nl{\tilde n_{1}} \def\nr{\tilde n_{2}}
\def\ns{N_{\rep 1}}
\def\nd{N_{\rep 2}}
\def\ninst{\cN}
\def\brk{\hfill\break}
\def\rep#1{\underline{#1}}
\def\one{{(1)}}
\def\NP{{({\rm NP})}}
\def\Re{{\rm Re\,}}

\def\B{V}
\def\BX{\B_X}
\def\Bb{\B_Y}
\def\hone{h^{(1,1)}}
\def\htwo{h^{(1,2)}}
\def\Y{Y}
\def\n{v}
%
%
\def\hepth#1{{\tt hep-th}/#1}
\ldf\CFG{N.~Seiberg, \nup 303 (1988) 286;\brk
S.~Cecotti, S.~Ferrara and L.~Girardello, Int.\ Mod.\ J.\ Phys.\
$\us{A4}$ (1989) 2475.}
\ldf\FHSV{S.~Ferrara, J.~Harvey, A.~Strominger and C.~Vafa,
\plt B361 (1995) 59, \hepth{9505162}.}
\ldf\GSW{M.~Green, J.~Schwarz and P.~West, \nup254 (1985) 327.}
\ldf\anosix{ J.~Erler, J.~Math.~Phys. $\us{35}$ (1994) 1819, 
\hepth{9304104};\brk
           J.~Schwarz, \hepth{9512053}.   }
\ldf\ANOMALIES{L.~Alvarez-Gaum\'e and E.~Witten, \nup234 (1984) 269;\brk
               A.~Salam and E.~Sezgin, Physica Scripta $\us{32}$ (1985) 283.}
\ldf\KKLMV{S.~Kachru, A.~Klemm, W.~Lerche, P.~Mayr and C.~Vafa,
\nup459 (1996) 537, \hepth{9508155}.}
\ldf\KV{S.~Kachru and C.~Vafa, \nup450 (1995) 69,
\hepth{9505105}.}
\ldf\CF{P.~Candelas and A.~Font, \hepth{9603170}.}
\ldf\KLT{V.~Kaplunovsky,  J.~Louis and S.~Theisen,
\plt B357 (1995) 71, \hepth{9506110}.}
\ldf\GH{O.~Ganor and A.~Hanany, \hepth{9602120}.}
\ldf\KLM{A.~Klemm, W.~Lerche and P.~Mayr, \plt B357 (1995) 313,
\hepth{9506112}.}
\ldf\MS{P.~Mayr and S.~Stieberger, \plt B355 (1995) 107,
\hepth{9504129}.}
\ldf\Duffetal{M.~Duff and J.~Lu, \nup357 (1991) 534;\brk
                  M.~Duff and R.~Khuri, \nup411 (1994) 473, \hepth{9305142};\brk
                  M.~Duff, \nup442 (1995) 47, \hepth{9501030}.
}
\ldf\SW{N.~Seiberg and E.~Witten, \nup426 (1994) 19;
\nup431 (1994) 484.}
\ldf\SWtwo{N.~Seiberg and E.~Witten, \hepth{9603003}.}
\ldf\BLPSSW{M.~Berkooz, R.~Leigh, J.~Polchinski, J.~Schwarz,
     N.~Seiberg and E.~Witten, \hepth{9605184}.}
\ldf\Sadov{V.~Sadov, \hepth{9606008}.}
\ldf\DMW{M.J.~Duff, R.~Minasian and E.~Witten,
\hepth{9601036}. }
\ldf\DLP{M.J.~Duff, H.~Lu and C.~Pope, \hepth{9603037}. }
\ldf\MV{D.R.~Morrison and  C.~Vafa, \hepth{9602114}; \hepth{9603161}.}
\ldf\FMS{S.~Ferrara, R.~Minasian and A.~Sagnotti, \hepth{9604097}.}
\ldf\FKM{S.~Ferrara, R.~Khuri and R.~Minasian, \hepth{9602102}.}
\ldf\CCLM{G.~Curio, \plt B368 (1996) 78, \hepth{9509146};\brk
G.L.~Cardoso, G.~Curio, D.~L\"ust and T.~Mohaupt,
\hepth{9603108}.}
\ldf\CCLMR{G.L.~Cardoso, G.~Curio, D.~L\"ust, T.~Mohaupt
and S.-J.~Rey, \nup 464 (1996) 18, \hepth{9603108}.}
\ldf\AG{P.~Aspinwall and M.~Gross, \hepth{9602118}.}
\ldf\AGtwo{P.~Aspinwall and M.~Gross, \hepth{9605131}.}
\ldf\BKKMSV{M.~Bershadsky, K.~Intriligator, S.~Kachru,
D.~Morrison, V.~Sadov and C.~Vafa,  \hepth{9605200} .}
\ldf\BCOV{M.~Bershadsky, S.~Cecotti, H.~Ooguri and C.~Vafa,
   \nup 405 (1993) 279, \hepth{9302103}; \cmp{165} (1994) 311, \hepth{9309140}.}
\ldf\AGM{P. Aspinwall, B. Greene and D. Morrison,
               \nup416 (1994) 414, \hepth{9309097}.}
\ldf\AFIQ{G.~Aldazabal, A.~Font, L.~Ib\'a\~nez and F.~Quevedo,
      \nup461 (1996) 85, \hepth{9510093}; \hepth{9602097}.}
\ldf\DKLL{B.~de Wit, V.~Kaplunovsky, J.~Louis and D.~L\"ust,
       \nup451 (1995) 53,  \hepth{9504006}.}
\ldf\EWa{E.~Witten,  \hepth{9507121 }; \nup460 (1996) 541,\brk
\hepth{9511030}.}
\ldf\EWb{E.~Witten, \hepth{9603150}.}
\ldf\EWc{E.~Witten, \nup403 (1993) 159, \hepth{9301042}.}
\ldf\HW{P.~Horava and E.~Witten, \nup460 (1996) 506,  \hepth{9510209};
               \hepth{9603142}.}
\ldf\AGNT{I.~Antoniadis, E.~Gava, K.~Narain and T.~Taylor,
              \nup455 (1995) 109,  \hepth{9507115};  \hepth{9604077}.}
\ldf\AFGNT{I.~Antoniadis, S.~Ferrara, E.~Gava, K.~Narain and T.~Taylor,
         \nup447 (1995) 35,  \hepth{9504034}.}
\ldf\west{M. Sohnius, K. Stelle and P. West, \plt B92 (1980) 123.}
\ldf\faux{P. Claus, B.~de Wit, M.~Faux, B. Kleijn, R. Sieblink and
          P. Termonia, \plt B373 (1996) 81, \hepth{9512143}.}
\ldf\HM{J.~Harvey and G.~Moore, \nup463 (1996) 315, \hepth{9510182}.}
\ldf\AL{P.S.~Aspinwall and J.~Louis, \plt B369 (1996) 233, \hepth{ 9510234}.}
\ldf\vafa{C.~Vafa,  \hepth{9602022}.}
\ldf\Batyrev{V. Batyrev, J. Alg. Geom. $\us 3$ (1994) 493;
                         Duke Math. J. $\us{69}$ (1993) 349.}
\ldf\toric{W. Fulton, {\it Introduction to Toric Varieties},
                      Princeton University Press, 1993; \brk
           M. Audin,  {\it The Topology of Torus Action on Symplectic
Manifolds}, \brk
                     Birkh\"auser 1991;\brk
           T. Oda,  {\it Convex Bodies and Algebraic Geometry}, 
                           Springer 1988;\brk
           V. Batyrev,  {\it Quantum Cohomology Rings of Toric Manifolds}.}
\ldf\HKT{S. Hosono, A. Klemm and S. Theisen,  {\it Lectures on Mirror
                     Symmetry},  in Springer LNP m20, 1994, \hepth{9403096}. }
\ldf\PF{S. Hosono, A. Klemm, S. Theisen and S.T. Yau,
                         \cmp 167 (1995) 301, \nup433 (1995) 501; \brk
    P. Berglund, S. Katz and A. Klemm, \nup456 (1995) 153.}
\ldf\sagnotti{A.~Sagnotti, \plt B294 (1992) 196.}
\ldf\BKKMI{P.~Berglund, S.~Katz, A.~Klemm and P.~Mayr, \hepth{9605154}.}
\ldf\BKKM{P.~Berglund, S.~Katz, A.~Klemm and P.~Mayr, in preparation
and talk given by P.~Berglund at the ITP, Santa Barbara, November 1995
and by A. Klemm at CERN, December 1995.}
\ldf\BKKMb{P.~Berglund, S.~Katz, A.~Klemm and P.~Mayr, \hepth{9605154}. }
%
%
%
\footline{\hss\tenrm--\folio\--\hss}
\def\pubnum{\hbox{LMU-TPW-96-15}\hbox{TAUP-2341-96}}
\titlepage
\vskip 1cm
\title{Non-Perturbative Properties   of Heterotic String Vacua
Compactified on ${K3\times T^2}$ $^{\diamond}$}
\author{J. Louis$^*$, J. Sonnenschein$^\dagger$, S. Theisen$^*$ and
            S. Yankielowicz$^\dagger$}
\vskip .7cm
\centerline{$^{*}$Sektion Physik der Universit\"at M\"unchen}
\centerline{Theresienstra\ss e 37, D - 80333 M\"unchen, FRG}
\medskip
\centerline{$^{\dagger}$ School of Physics and Astronomy}
\centerline{Beverly and Raimond Sackler Faculty of Exact Sciences}
\centerline{Tel Aviv University}
\centerline{Ramat-Aviv, Tel-Aviv 69978, Israel}
\vskip .5in

\footnote{}{$^{\diamond}$ Work supported by GIF - the German-Israeli 
Foundation for Scientific Research.

\ni
$^*$ Work supported by the European Community Research Programme 
under contract SCI-CT92-0789.

\ni
$^\dagger$ Work supported in part by Israel Academy of Science.}

\abstract{
Using the heterotic--type II  duality of $N=2$ string vacua
in four space-time dimensions we study  non-perturbative
couplings  of toroidally compactified six-dimensional heterotic vacua.
In particular, the heterotic--heterotic $S$-duality and
the Coulomb branch of  tensor multiplets observed in
six dimensions are studied from a four-dimensional point of view.
We explicitly compute the couplings of the vector multiplets of several
type II vacua and investigate the implications for their heterotic duals.}

\vskip 2cm

\endpage

%
\chapter{Introduction}
During the past year it has become clear that some string theories
and their vacuum states are connected in an intricate fashion.
The various interrelations and their physical
implications strongly depend on the number of space-time
dimensions and the amount of supersymmetry
of the string vacua under  consideration.
Recently, heterotic vacua in six dimensions ($d=6$)
with  minimal ($N=1$) supersymmetry have been under active
investigation. Such vacua  can be constructed
in string perturbation theory by compactifying
the ten-dimensional heterotic string on a $K3$ surface.
The massless spectrum is strongly  constrained
by the  cancellation of gauge and gravitational anomalies and
the gauge bundle is required to have non-trivial instanton
numbers \nrf{\ANOMALIES\GSW\anosix}\refs{\ANOMALIES-\anosix}.

The gauge bundle becomes singular when an instanton
shrinks to  zero size \EWa.
This  singularity  occurs at arbitrarily weak string coupling
but nevertheless cannot be seen in string perturbation theory;
rather it appears in regions of   the  moduli space
where the conformal field theory description of a  string  vacuum breaks down.
For $SO(32)$  heterotic vacua the singularity is caused by non-perturbative
gauge fields which become massless at the locus (in moduli space)
of the shrinking instanton and which
enhance the rank of the perturbative gauge group beyond the bound
implied by the central charge \EWa.
On the other hand  in a generic  $E_8\times E_8$ vacuum
it is believed that at the singularity 
a non-critical  string becomes
tensionless \nrf{\GH\SWtwo\DLP}\refs{\GH-\DLP}.
This singularity signals the transition to a non-perturbative phase
with  extra tensor multiplets. (In perturbative heterotic vacua there is always
exactly one tensor multiplet.)
In $d=6$ a tensor
multiplet contains an anti-selfdual  antisymmetric tensor and a real scalar
field as bosonic components.
Therefore, a new non-perturbative `Coulomb-branch'  parameterized
by the vacuum expectation values of the additional scalars  exists;
this branch is invisible in string perturbation theory.

The non-perturbative physics of the heterotic vacua is captured by
M-theory compactified on $K3\times S^1/\ZZ_2$ \HW\ and/or by
F-theory compactified on elliptically fibered Calabi--Yau threefolds
\nrf{\vafa\MV\EWb}\refs{\vafa-\EWb}.
In M-theory there is an $E_8$ gauge factor associated with each
of the two nine-branes which sit at the fixed points of $S^1/\ZZ_2$
and there are dynamical  five-branes with massless tensor multiplets.
In this picture the  transition to the  non-perturbative Coulomb  branch
corresponds to a five-brane leaving one of the nine-branes
and the tensionless string emerges from a collapsed two-brane that connects the
five-brane to the nine-brane \refs{\DMW\GH, \SWtwo}.
The string is an effective description
of the two-brane when the five and the nine-branes are close
to each other. Its tension is linearly dependent on the separation and
when it vanishes one gets a tensionless string.
In F-theory the same transition is described by
blowing up the base manifold of the elliptically fibred Calabi--Yau threefold
\refs{\MV, \EWb}.

Apart from the weak coupling singularities just discussed
there is generically also a strong coupling singularity
where the normalization of the gauge kinetic terms
turns negative \DMW. This singularity is believed to
result from a non-critical string becoming tensionless
with its tension controlled by the dilaton \SWtwo.
For heterotic $E_8\times E_8$ vacua with equal instanton number 
in each group factor the strong coupling singularity is absent
and a strong-weak or S-duality is conjectured to hold \DMW.
Only in this case can a five-brane be consistently wrapped around
the $K3$. This results in a new  string which is identified as the dual
heterotic string. The dual heterotic vacuum
has the inverse string coupling constant, the antisymmetric tensor is replaced
by its dual,  the moduli space of the hypermultiplets is mapped non-trivially
onto itself and finally perturbative and non-perturbative
gauge fields are interchanged.
The existence of non-perturbative gauge fields
is a prerequisite for the heterotic--heterotic duality.
Recently it has been shown  that their appearance
in $E_8\times E_8$  vacua can be understood via the $T$-dual
${Spin(32)/\ZZ_2}$ vacuum whose
small instantons are responsible for the non-perturbative gauge 
symmetries \BLPSSW.
Further support for the validity of the  heterotic-heterotic duality
has been accumulated in  refs.~\refs{\AFIQ,\MV,\AG}.

The special properties of the six-dimensional vacua
can also be observed in toroidally compactified
vacua with four space-time dimensions
and $N=2$ supersymmetry. In $d=4$  the heterotic--heterotic duality is no longer
a strong--weak coupling duality
but rather involves the exchange of the four-dimensional
dilaton $S$ with the radial modulus
$T$ of the two-torus \nrf{\Duffetal\KLM\CCLMR}\refs{\Duffetal-\CCLMR, \DMW}.
On the other hand the map among  the hypermoduli as well as
the interchange of perturbative with non-perturbative gauge fields
continues to hold in the compactified vacua.
Similarly, the tensor multiplets of the six-dimensional vacua
turn into vector-tensor multiplets in $d=4$ which are dual
to vector multiplets \nrf{\west\DKLL\faux}\refs{\west-\faux}.
Thus in $d=4$ the non-perturbative Coulomb branch of the tensor multiplets
turns into a non-perturbative Coulomb branch in the four-dimensional  moduli
space of the vector multiplets.

In $d=4$ the $N=2$ heterotic vacua are believed to be
non-perturbatively equivalent to $N=2$ vacua of the type II string
\refs{\KV,\FHSV}.
In particular, the non-perturbative physics
of the gauge sector in the heterotic string is captured
by a weakly coupled type II vacuum and thus can be seen
in type II perturbation theory. This implies
that the properties of the non-perturbative
gauge fields (including the exchange symmetry with the perturbative
gauge fields) as well as the Coulomb branch of the tensor
multiplets  should  be visible in the appropriate type II vacua.

In this paper we focus on a number
of explicit $d=4$ heterotic vacua  and their
dual type II description. We compute the
couplings of the vector multiplets and display consequences of
the (non-perturbative) properties of the $d=6$ heterotic vacua.
The organization of the material is as follows.
In section~2.1 we briefly recall the properties of $N=1$ heterotic vacua
in $d=6$. In 2.2 we discuss the toroidal compactification of these
vacua and the  specific structure of their gauge couplings.
Section~3 is devoted to the construction (3.1)
and the computation of the couplings (3.2 -- 3.4) of the dual type II vacua.
The physical implications for the heterotic vacua are  discussed as we go along.

%
%
\chapter{The heterotic string}
\section{$E_8\times E_8$ heterotic vacua in $d=6$}

In this section we briefly recall the main features
of heterotic vacua in six dimensions.
Their spectra are constrained by gravitational
and gauge anomaly cancellation.
In particular, the vanishing of the $\tr R^4$ term demands \ANOMALIES
$$
\nh - \nv + 29\, \nt = 273\ ,
\eqn\Rfour
$$
where $\nh,\nv,\nt$ counts the number of hyper, vector and
anti-selfdual tensor multiplets, respectively.
The remaining anomaly eight form $I_8$ has to be cancelled by appropriate
Chern--Simons interactions of the antisymmetric tensor fields
\nrf{\sagnotti}\refs{\GSW, \sagnotti}.

Perturbative  heterotic vacua in $d=6$  are obtained by compactifying
the ten-dimensional heterotic string on a $K3$ surface.
In this case the massless spectrum
contains one tensor field (\ie\ $\nt=1$)\foot{There is an anti-selfdual
tensor in the tensor multiplet and a selfdual tensor in the gravitational
multiplet. They combine to one unconstrained antisymmetric tensor $B$.},
$I_8$ factorizes $I_8=X_4\cdot \tilde X_4$
and the field strength $H$ of the antisymmetric tensor
obeys the Bianchi identity $dH= X_4$. In order to ensure a globally
defined three form $H$ on the compact $K3$
the integral $\int_{K3} dH$ has to vanish.
For $E_8\times E_8$ vacua where
$X_4  = \tr R\wedge R - \sum_{a=1,2}  v_a \tr (F\wedge F)_a $,
$\tilde X_4 = \tr R\wedge R -  \sum_{a=1,2} \tilde v_a \tr (F\wedge F)_a$,
(the constants $v_{a}\,(\tilde v_{a})$ are given in ref.~\anosix)
the gauge bundle
has to have non-trivial instanton configurations which  obey
$$
n_1 + n_2 = 24\ .
\eqn\inumber
$$
Here $n_1$ and $n_2$ are the instanton numbers of the two
$E_8$ factors and $24$ is the Euler number of $K3$.

For an arbitrary gauge group $G$ the moduli space of instantons on $K3$
is a quaternionic manifold of (quaternionic) dimension
$$
\ninst_{n} [G] = n\, h -  {\rm dim}(G)\ ,
\eqn\diminst
$$
where $n$ is the instanton number
and $h$ the dual Coxeter number of $G$.
The gauge bundle becomes singular in the limit of a zero size instanton.
In $E_8\times E_8$ vacua this phase transition is associated 
with the generation of
additional massless tensor multiplets which cannot be seen in string
perturbation theory. Indeed, from eq.~\diminst\ we learn that by shrinking an
$E_8$ instantons the dimension of the moduli space drops by $30-1=29$ 
where the one extra modulus  parametrizes the location of the small instanton.
$29$  is precisely the number of hypermultiplets which can be traded
with a tensor multiplet while leaving the constraint \Rfour\  intact.
If additional tensor multiplets are present the constraint \inumber\ has to be
modified according to
$$
n_1 + n_2 + \nt = 25\ ,
\eqn\inumbernp
$$
and 
$I_8$ no longer factorizes but splits into a sum of two terms \SWtwo
$$
I_8 (n_1,n_2) = [\coeff12\,A_0- A_1]
\cdot[\coeff{n_1-8}{4}\,A_0-\coeff{n_1-12}{2}\,A_1]
+[\coeff12\,A_0-A_2]
\cdot[\coeff{n_2-8}{4}\,A_0 -\coeff{n_2-12}{2}\,A_2]\ ,
\eqn\Ieight
$$
where we abbreviated $A_0\equiv \tr R\wedge R,
A_1 \equiv v_1 \tr(F\wedge F)_1, A_2 \equiv v_2 \tr(F\wedge F)_2$.

In the perturbative limit ($\nt=1$) eq.~\Ieight\   factorizes
and the anomaly is cancelled by a (conventional) Green--Schwarz
mechanism where the field strength of the antisymmetric tensor is defined by
$H=dB +\omega_{\rm L} -\sum_{a=1,2}  v_a\,\omega_{\rm YM}^a$
($\omega_{\rm L}(\omega_{\rm YM})$ are the Lorentz (Yang--Mills)
Chern--Simons terms) such that $dH=X_4$.
In the
generic case with more than one tensor multiplet $I_8$ does  not factorize.
A generalized  Green--Schwarz  mechanism is necessary
where the additional
tensor fields are also required to have  appropriate Chern--Simons couplings
to the gauge and gravitational fields \refs{\sagnotti, \SWtwo, \FMS,\Sadov}.
These couplings become apparent when one rewrites \Ieight\ as
$$
I_8 (n_1,n_2)=I_8(12-k,12+k)
-\coeff{\nl}{2}\,[\coeff12\, A_0-A_1]^2
-\coeff{\nr}{2}\,[\coeff12\, A_0 - A_2]^2\,
\eqn\Isum
$$
where $\nl (\nr)$ is the number of small instantons in the first (second) $E_8$
factor and
$n_1=12-k-\nl, n_2=12+k-\nr, \nt=1+\nl+\nr$ holds.
Eq.~\Isum\  reveals that the extra terms are two  perfect squares
each  of which only depends on one of the two $E_8$ factors \HW.
Such contributions to the anomaly can be  cancelled by
Chern-Simons interactions of  the $(\nl+\nr)$
additional anti-selfdual tensor fields \sagnotti.
However, the fact that each of the extra terms in eq.~\Isum\ only depends
on one of the $E_8$ factors implies that also the Chern--Simons
terms in the corresponding tensor field only depends on that same
$E_8$ factor. Note that specifying $n_1,n_2$ does not uniquely determine
$k$ and $\tilde n_1,\tilde n_2$ or, in other words, there is an ambiguity in
assigning the Chern-Simons terms of the extra tensors.

The scalars of the $\tilde n_1+\tilde n_2$ tensor multiplets parametrize a
non-perturbative branch of the moduli space which opens up on a
subspace of the hypermultiplet moduli space
corresponding to a small instanton.
The transition to the new branch can be observed in
M-theory compactified on $K3\times S^1/\ZZ_2$;
it corresponds to a five-brane that has been detached from the nine-brane
and which carries the additional tensor. Furthermore,
when the five-brane is  `swallowed' by the other nine-brane
a second transition occurs to a heterotic vacuum
with  instanton numbers  $(n_1-1, n_2+1)$.
Note that the Coulomb branch on which we have an extra tensor does not 
seem to have a direct geometrical interpretation from a $d=10$ point of view.

In the F-theory description of the heterotic vacua one has to choose
elliptic  Calabi--Yau threefolds $Y$ as compact manifolds
\refs{\vafa, \MV}.
There is then a (regular, connected)
holomorphic map $Y\to B$ such that the generic fiber $Y_b\,(b\in B)$ is
a smooth elliptic curve.
The number of tensor multiplets is directly related
to the number of $(1,1)$-forms on the base $B$  via
$$
\nt = \hone (B) - 1\ .
\eqn\Fcount
$$
In this context
the perturbative heterotic vacua with instanton numbers
$(12-k,12+k)$  are identified with elliptic fibrations over
the Hirzebruch surfaces $\IF_k$.
The $\IF_k$ have
$h^{(1,1)}=2$ (\ie\ $\nt=1$) consistent with their perturbative
interpretation but they  can be blown up to give additional $(1,1)$-forms
which in terms of the heterotic vacuum correspond  to
new tensor multiplets. The transitions  between the perturbative
and non-perturbative
heterotic vacua are thus seen as transitions among elliptically fibered
Calabi--Yau threefolds with blown up and blown down Hirzebruch 
surfaces as their base.
In particular the transition $(n_1, n_2) \to (n_1-1, n_2+1) $ is
identified with the transition $\IF_k \to \IF_{k+1}$.

For $n_1,n_2 > 9$ the instantons generically break the gauge
group completely and one is  left with only
tensor multiplets and gauge
neutral hypermultiplets.
The number of hypermultiplets is determined by the dimension
of the instanton moduli space (eq.~\diminst) together with
20 additional  quaternionic moduli of the $K3$ surface
and $(\nt-1)$ hypermultiplets which parameterize the location of the
small instantons.
Therefore the total number of hypermultiplets is found to be
$$
\nh =  20 + (\nt-1) + \ninst_{n_1}[E_8] + \ninst_{n_2}[E_8] 
= 273-29\, \nt\ ,
\eqn\idim
$$
where the last equation uses \diminst, \inumbernp\ and, as required,
the constraint \Rfour\ is satisfied.

If the instantons are embedded in a subgroup $H$ of $E_8\times E_8$
the heterotic vacuum is left with some gauge symmetry,
charged matter multiplets and neutral moduli.
The decomposition of the adjoint representation of $E_8$ 
into the representations $\rep{h_i}$ of $H$
and the representations $\rep{g_i}$ of the commutant of $H$ --
$\rep{248} = \sum_i (\rep{g_i}, \rep{h_i})$ -- determines
the number of charged hypermultiplet $N_{\rep{g_i}}$
according to \GSW
$$
N_{\rep{g_i}} = \coeff12 l(\rep{h_i}) \,  n  - {\rm dim}\,  \rep{h_i}\ .
\eqn\nmatter
$$
($l(\rep{h_i})$ is the index of the representation $\rep{h_i}$.)

For example, embedding the instantons into $E_8\times E_7$ leaves
an unbroken gauge group $SU(2)$ with $\ns$ singlets and
$\nd$ doublets
$$
\eqalign{
\ns = &\
20 + (\nt-1) + \ninst_{n_1}[E_8] + \ninst_{n_2}[E_7]
= 29\, n_1 +17\, n_2  - 337\ ,\cr
\nd =&\ 6\,  n_2 - 56 \ .
}
\eqn\idimg
$$
The total number of hypermultiplets is
$\nh = \ns + 2 \nd = 273+3 - 29\nt$
consistent with $\Rfour$.
The difference in the number of singlets compared to \idim\
is $ \ninst_{n_2}[E_8] - \ninst_{n_2}[E_7] = 12\, n_2 -115$
or, in other words, one has to  tune $12\, n_2 -115$ hypermultiplets
to open up an $SU(2)$ gauge symmetry.

For future reference
let us record a few more spectra:
$$
\eqalign{
SU(2)_1\times SU(2)_2:\qquad
\ns = &\
20 + (\nt-1) + \ninst_{n_1}[E_7] + \ninst_{n_2}[E_7] \cr
= &\ 17\, (n_1 + n_2)  - 222\ ,\cr
N_{(\rep{2},\rep 1)}&\  =  6\,  n_1 - 56 \ , \qquad
N_{(\rep 1,\rep{2})} =  6\,  n_2 - 56 \ , \cr
}
\eqn\suot
$$
where the two $SU(2)$'s arise from different  $E_8$ factors.
$$
\eqalign{
SU(2)_1\times SU(2)_1:\qquad
\ns = &\
20 + (\nt-1) + \ninst_{n_1}[SO(12)] + \ninst_{n_2}[E_8] \cr
=&\  9\, n_1 + 29\, n_2   - 270\cr
N_{(\rep{2},\rep 1)}&\ =  4\,  n_1 - 32 \ , \qquad
N_{(\rep 1,\rep{2})} =  4\,  n_1 - 32 \ ,  \cr
N_{(\rep 2,\rep{2})}&\  =   n_1 - 12  \ ,
}
\eqn\suoo
$$
here the two $SU(2)$'s arise from the same $E_8$ factor.\foot{Note that
for $n_1<12$, $N_{(\rep 2,\rep 2)}$ is negative; the chirality
assignments of the spinors in the various $d=6$ supermultiplets render
this vacuum inconsistent. One arrives at the same conclusion using the
Higgs mechanism.}
$$
\eqalign{
E_7\times E_7:\qquad
\ns = &\
20 + (\nt-1) + \ninst_{n_1}[SU(2)] + \ninst_{n_2}[SU(2)] \cr
=&\  n_1 + n_2   +38\ , \cr
N_{(\rep{56},\rep 1)}&\ =\  \coeff12  (n_1 - 4) \ , \qquad
N_{(\rep 1,\rep{56})} =  \coeff12  (n_2 - 4) \ .
}
\eqn\eses
$$
All spectra obey the constraint \Rfour.

In \eses\  the  instantons are embedded
into $SU(2)_1\times SU(2)_2$ and the gauge symmetry is $E_7\times E_7$.
One can use a standard Higgs mechanism by  giving
appropriate  vacuum expectation values to the
$(1,\rep{56})$ and $(\rep{56},1)$
multiplets to obtain the same spectra \idimg--\suoo\
of massless modes.\foot{For
example, breaking $E_7\times E_7\to SU(2)$
requires a decomposition of  the $E_7$ under
its  maximal subgroup containing $SU(2)$ which is $SO(12) \times SU(2)$.
The relevant representations decompose according to
$\rep{133} \to (\rep{1},\rep{3}) + (\rep{66},\rep{1})+(\rep{32'},\rep{2})$,
$\rep{56} \to (\rep{32},\rep{1}) + (\rep{12},\rep{2})$.
A VEV of the $(\rep{32},\rep{1})$ breaks $E_7\to SU(2)$ with a spectrum
of $\left(16 n_2 - 130\right)$ singlets and
$\left(6 n_2 - 56\right)$ doublets.
Together with Higgsing the second $E_7$ completely one recovers
the same spectrum as in eq.~\idimg.
}

In perturbative vacua the normalization of the gauge kinetic terms
is fixed by supersymmetry to be\foot{In the Einstein frame this corresponds to
$
{\cal L}\sim\sqrt{G_E}
\sum_a\, (v_a e^{-\Phi/2} + \tilde v_a e^{\Phi/2})\ \tr_a F^2
$
with $G_E$ being the metric in the Einstein frame.}
$$
{\cal L} \sim\  \sqrt{G} \,
\sum_{a=1,2}\, (v_a e^{-\Phi} + \tilde v_a)\ \tr_a F^2\ ,
\eqn\gauge
$$
where $\Phi$ is the six-dimensional dilaton and $G$ the metric
in the string frame. 
This indicates that there is a strong coupling singularity
whenever $e^{-\Phi} = -\tilde v/v=|k|/2$.
It is believed that this singularity is caused by a string whose
tension is set by the dilaton and which approaches zero at
the critical value of the dilaton \refs{\SWtwo,\DLP}.
For $n_1=n_2=12$, i.e. $k=0$ there is no strong coupling singularity
and it  takes the same number of parameters ($12\cdot12-115=29$)
to open up an $SU(2)$ gauge group as is needed to shrink
an $E_8$ or $SO(32)$ instanton.\foot{A shrinking $E_8$ or $SO(32)$  instanton
always  requires  tuning 29 hypermultiplets
but the un--Higgsing of an $SU(2)$ from a completely Higgsed phase takes
$12\,n-115$ parameters which only coincide for $n=12$.}
A small $E_8$ instanton always leads to a tensionless string but
in $(12,12)$ vacua of the $E_8\times E_8$ heterotic string
small ${Spin(32)/\ZZ_2}$ instantons can exist which induce non-perturbative
gauge fields.
This is possible due to $T$ duality between the $E_8\times E_8$ heterotic and
the
${Spin(32)/\ZZ_2}$ Type I string \BLPSSW.
Indeed, in ref. \DMW\ a heterotic--heterotic
self-duality of the $(12,12)$ vacua was conjectured. One  replaces
$$
\Phi \to  - \Phi\ , \qquad
G \to  e^{-\Phi} G\ , \qquad
H \to  e^{-\Phi} *H \ ,
\eqn\map
$$
and in addition exchanges
perturbative and non-perturbative gauge fields.
As we just saw the perturbative and non-perturbative gauge
symmetry appears on subspaces of the hypermultiplet moduli space
which have the same dimension. However, these subspaces
are not  identical and therefore the exchange of perturbative
with non-perturbative gauge fields necessarily requires
a non-trivial map between the hypermultiplets.
Let us also remark that  S-duality is consistent with the absence of a strong
coupling singularity since perturbatively we know that $v_\alpha>0$ and using
duality this implies that also $\tilde v_\alpha\geq0$.
{}From the M-theory point of view the duality holds only in the instanton
symmetric case since only then one has an additional string which arises from
wrapping a five-brane over $K3$.

In this paper our main interest are the four-dimensional consequences
of the physical phenomena just described. Therefore, let us now turn to
toroidally compactified heterotic vacua.

%
\section{Heterotic vacua in $d=4$}
Compactification of the $d=6,N=1$ heterotic vacua on a two-torus $T^2$
yields  four-dimensional  vacua with $N=2$ supersymmetry.
A hypermultiplet is untouched in the compactification
while a vector multiplet gains a complex scalar
in the adjoint representation of the gauge group.
The  scalars $C^i, i=1,\ldots,{\rm rank}(G)$
in a Cartan subalgebra of $G$ are flat directions of the effective
potential and at generic values in their field space
the gauge group is broken to  $[U(1)]^{{\rm rank}(G)}$.
Thus, in $d=4$ there is a Coulomb branch in the moduli space
parametrized by the vacuum expectation values of $C^i$'s.
(This branch does not exist in the six-dimensional vacua
since the $d=6$ vector multiplets
do not contain a scalar degree of freedom.)
Furthermore, in  toroidally compactified vacua
there always are two additional Abelian vector multiplets
-- denoted by $T$ and $U$ --
which contain the Kaluza--Klein gauge bosons of the torus and
the corresponding toroidal moduli.\foot{
A third  vector turns into the graviphoton which resides in the
gravitational multiplet.}

A dimensionally reduced tensor multiplet turns into a
vector--tensor multiplet \refs{\west-\faux}
which contains an antisymmetric tensor,
a vector and a real scalar as bosonic components. In $d=4$
an antisymmetric tensor is dual to a scalar and hence a vector--tensor multiplet
can be dualized to give another vector multiplet.
In perturbative heterotic vacua
there is exactly one such  multiplet -- denoted by $S$ --
which contains the four dimensional dilaton.
However, as we saw in the previous section, additional
vector-tensor multiplets can appear and
we  denote their dual vector multiplets collectively by $\B$.
Similarly, non-perturbative vector multiplets can arise on singular
subspaces of the hypermultiplet moduli space.
These multiplets also have a Coulomb branch parameterized by
their scalars $C^{\prime i'}$
which are in the Cartan subalgebra of the non-perturbative gauge group.

At the two-derivative level the couplings of the vector multiplets are  encoded
in a holomorphic prepotential $\Fhet$.
This prepotential can be computed in string perturbation theory
where it receives a contribution at the tree level and at one-loop
but not beyond. For heterotic vacua which arise as $T^2$
compactifications one finds \DKLL\foot{Here 
we have slightly changed the conventions compared to ref.~\DKLL\
in order to simplify the correspondence with the dual type II vacua
in the next section. In particular, we rescaled $\Fhet$ by an
overall $-4\pi$ along with a scaling of $S$ by $4\pi$.}
$$
\Fhet = S(TU - \sum_i  C^i C^i) + \Fhet^\one (T,U,C)
                                  + \Fhet^\NP (e^{-2\pi S},T,U,C,C',\B)\ ,
\eqn\Fhetet
$$
where the first term is the tree level result,
$\Fhet^\one$ is the (dilaton-independent) one-loop contribution
and $\Fhet^\NP$ summarizes the possible non-perturbative corrections.
In this parametrization a large $S$ is the
 weak coupling (perturbative) expansion parameter.
$\Fhet^\one (T,U,C)$ generically depends on the
specific properties of the heterotic vacuum
under consideration. However, precisely when such vacua arise as
toroidal compactifications the $T$ and $U$ dependence
can be computed \nrf{\AFGNT\MS\HM}\refs{\DKLL, \AFGNT, \MS, \HM}.
This is largely due to the fact that there is a perturbative
symmetry $SL(2,{\ZZ})_T \times SL(2,{\ZZ})_U$ acting on
the moduli $T$ and $U$\foot{Here and throughout the paper
we use the same symbol for a
vector multiplet and its  scalar component.}
which strongly  constrains the one-loop
corrections of $\Fhet$.
One finds that the third derivatives of $\Fhet^\one$
with respect to $T$ and $U$ as well as
the second derivative with respect to $C^i$
have to be  specific
modular forms  of $SL(2,{\ZZ})_T \times SL(2,{\ZZ})_U$;
they can be integrated to give $\Fhet^\one$ \HM.
For our present purpose we only need the leading contribution
of $\Fhet^\one$ in the large $T, U$ limit which is 
$$
\Fhet^\one = P^\one_3(T,U,C^i)  + \ldots\ .
\eqn\Fhetint
$$
Here $P^\one_3$ is a cubic polynomial of its arguments and the
ellipses stand for subleading terms.
$P^\one_3$ is not uniquely
defined since in perturbation theory
the dilaton $S$ can be shifted $S\to S + \a T + \b U$ where
$\a,\b$ are arbitrary complex constants.
Such a shift in the first term of eq.~\Fhetet\
redefines $P^\one_3$ by a cubic polynomial  of the form
$P^\one_3 \to P^\one_3 + \a\, T^2 U + \b\, TU^2$ but  no  cubic terms
$T^3$ or $U^3$ can be generated.
Such  terms in $P^\one_3$ have an invariant
meaning and have  been computed in ref.~\HM.
However, there is a further complication due to the fact
that $\Fhet^\one$ has a singularity at $T=U$ (mod $SL(2,{\bf Z})$).
On this subspace of the moduli space additional
gauge bosons become massless and the toroidal
gauge group $U(1)_T\times U(1)_U$ is enhanced
to $SU(2)\times U(1)$.\foot{At $T=U=1$ and $T=U=e^{i\pi/6}$
there is a further enhancement to $SU(2)^2$ and $SU(3)$, respectively.}
The cubic terms in $P_3^\one$ are sensitive to the region
(the `Weyl chamber') where the computation is done.
Choosing a definition of the dilaton such that  $P^\one_3$ contains no
terms $T^2 U$ and $TU^2$
one finds \refs{\HM,\DKLL,\AFIQ}
$$
\eqalign{
P^\one_3 =&\  \coeff13 U^3 \
-\  \coeff1{12}\big( (b-12)\, T + b \, U \big)\, C^iC^i
  \qquad
    {\rm for}\ \Re T > \Re U \ , \cr
 P^\one_3 =&\ \coeff13 T^3\
-\  \coeff1{12}\big( (b-12)\, U + b\, T \big)\, C^iC^i
    \qquad
    {\rm for}\ \Re U > \Re T  \ .
}
\eqn\Presult
$$
Here $b=N_{\rep{r}}\,l(\rep{r})-l(\rep{\rm ad})$ is the coefficient of the
beta function of $G$. 
The prefactor of the first term has been computed for vacua with only
$S,T,U$. In section~3  we  observe that in the dual type II vacua
the same coefficient seems to be $(9-\nt)/24$  (in a basis to be 
specified below) but we have no independent confirmation  from a
heterotic consideration. Similarly the coefficients of $T C^i C^i$ and
$U C^i C^i$ are modified in the presence of $V C^i C^i$ couplings.

The non-perturbative corrections $\Fhet^\NP$ in eq.~\Fhetet\
summarize  the space-time instanton corrections to $\Fhet$.
Such contributions are supressed by $e^{-2\pi S}$ and therefore
vanish in the weak coupling $\Re S \to \infty$ limit.
However, as we already discussed, there can be additional
vector multiplets $C'$ and/or dualized vector-tensor multiplets $\B$
which are
of non-perturbative origin and do not have the canonical couplings to
the dilaton. In our notation we have included their entire couplings
into $\Fhet^\NP$ indicating that their contribution to the prepotential
cannot be computed in heterotic perturbation theory.
With this convention, $\Fhet^\NP$ does not vanish in the large $S$ limit
but rather obeys
$$
\Fhet^\NP \to P_3^\NP(T,U,C,C',\B) \qquad {\rm for}\ S\to\infty\ ,
\eqn\FNPgeneric
$$
where $P_3^\NP$ is a cubic polynomial of its arguments
but it does not depend on the dilaton $S$.
The couplings of 
$\B$ are constrained purely from supergravity considerations.
As we saw in section~2.1  antisymmetric tensor fields generically
have Lorentz and Yang--Mills Chern--Simons couplings.
Here  we need to distinguish two different types of dualized vector-tensor
multiplets. If the antisymmetric tensor only couples to Lorentz 
and Yang--Mills Chern--Simons
terms of the graviphoton and its own  Abelian vector partner,
the dual vector multiplet -- $\BX$ in the following --
only appears as $\BX^3$ in $P_3^\NP$ \faux.
On the other hand, if an antisymmetric tensor couples to Chern--Simons
terms of other, in particular  non-Abelian gauge fields,
then the dual vector multiplet -- which we denote $\Bb$ --
can never appear cubic but at most quadratic in  $P_3^\NP$ \FMS.
Furthermore, the coupling of $\Bb$ to the vector multiplets present
in the six-dimensional vacuum $C,C'$ is always linear.
A more detailed analysis can be found in ref.~\FMS\ but for
our purpose we record that for $\Re T > \Re U$ one has
$$
P_3^\NP =\coeff16 \BX^3 + \Bb \Big( \sum_i \gamma_i C^i C^i
                + \sum_{i'} \gamma'_{i'} C'^{i'} C'^{i'} \Big)
 -\coeff12 U \Bb^2 \, + \, \tilde P_3^{\NP}(U,T,C')\ ,
\eqn\purecube
$$
where $\tilde P_3^\NP$ is a model dependent cubic polynomial and 
$\gamma_{i}, \gamma'_{i'}$ are constants directly related to the
Chern--Simons couplings of the dual tensor field.
In particular one has $\gamma_i (\gamma'_{i'})=0$ if the tensor  does
not couple to the Chern--Simons term of $C^i (C'^{i'})$.
In section~2.1 we learned that the tensor fields in $d=6$
only couple to one of the $E_8$ factors but not the other.
(For $\Re U > \Re T$ the roles of $T$ and $U$ are interchanged in eq.~\purecube.)

The prepotential $\Fhet$ encodes the
couplings of the gauge fields at the two derivative level.
Certain higher derivative couplings of vector multiplets are also encoded
in holomorphic sections $F_g$ whose weak coupling behaviour is known.
In particular the coupling to $R\wedge R$  resides in $F_1$
which in the large $S$ limit  obeys \refs{\KLT, \AGNT}
$$
F_1 = 24\,  S + P_1(T,U,\B,C,C') + \ldots\ .
\eqn\Fonelarge
$$
$P_1$ is a linear polynomial in its variables and the ellipses
stand for terms which vanish as $S\to\infty$.
$P_1$ depends on the specific vacuum under consideration but
from eq.~\Isum, taking into account the normalization of the dilaton in
eq.~\Fonelarge, we can infer the dependence on the
antisymmetric tensors to be
$$
P_1 = - 12\,  \Bb + \ldots 
\eqn\twelve
$$
(the choice of sign is a matter of convention and correlated with the sign 
of $\gamma_i$ in eq. \purecube).
In perturbative heterotic vacua also the $T$ and $U$ dependence of 
$P_1$ is known to be $24 T+44 U$ \CCLM; the coefficients change if 
$V_Y$'s are present in the spectrum.

As the final point of this section let us note that the
heterotic--heterotic duality
discussed in section~2.1 has its traces in $d=4$.
However, it is no longer a strong--weak coupling duality but rather
an exchange symmetry between the four-dimensional dilaton
and the radial K\"ahler modulus of the two-torus.
The four-dimensional dilaton which
coincides with the leading (tree-level) term
of the perturbative gauge couplings is the real part of
the complex scalar $S$.
By dimensional reduction one finds the relation with the six-dimensional
dilaton $\Phi$ via the couplings \gauge
$$
\Re S = r^2 e^{-\Phi}\ ,
\eqn\foursixrel
$$
where $r$ is the radius of the two-torus.
On the other hand the modulus $T$ which paramterizes the volume
of the two-torus is
$$
\Re T = r^2 \ .
\eqn\Trel
$$
Using \gauge, \map, \foursixrel\ and \Trel\
it is straightforward to show  that the $d=6$  heterotic--heterotic duality
turns into the exchange $S\leftrightarrow T$ in $d=4$ together with a
map of the hypermultiplets and
the exchange of perturbative and non-perturbative gauge fields
\refs{\Duffetal, \DMW}.
In particular, these properties should be manifest in the heterotic prepotential
$\Fhet$ given in \Fhetet.
Within a purely perturbative definition of the heterotic string these
features can neither be observed nor computed.  However,
it  is believed that at least a subclass of  heterotic $K3\times T^2$
compactifications are non-perturbatively equivalent
to Calabi--Yau compactifications of the type IIA string \refs{\KV, \FHSV}.
With this duality at our disposal it should be possible to
observe the non-perturbative properties of the heterotic
string which we discussed in this section.
Therefore, we now turn to a discussion of the dual type II
vacua.

\chapter{ The type  IIA string compactified on Calabi--Yau manifolds}

String vacua which result from compactifying the type II  string on a
Calabi--Yau threefold $\Y$ also have $N=2$ supersymmetry 
in four space-time
dimensions. The dilaton and the antisymmetric tensor  together with
two universal scalar degrees of freedom from the Ramond--Ramond sector
form an $N=2$ tensor multiplet, which 
is different from the vector-tensor multiplet discussed
previously in that it contains no vector field.  Upon dualizing
the antisymmetric tensor this multiplet turns into a hypermultiplet and
as a consequence the dilaton in type II vacua always lives 
in this universal  hypermultiplet.
Further hypermultiplets arise in type IIA vacua
from the $(1,2)$ moduli  of  the  Calabi--Yau
manifold while  the $(1,1)$ forms are in one-to-one
correspondence with Abelian vector multiplets on the Coulomb branch \CFG.
Altogether we  have
$$
\nh = \htwo(\Y) + 1\ , \qquad \nv = \hone (\Y)\ .
\eqn\dimtt
$$

Locally the moduli space between hyper and vector
multiplets factorizes  and thus the classical moduli space of the vector
multiplets is exact in type II vacua. (The same argument shows that the
moduli space of the hypermultiplets is exact in heterotic vacua.)
The equivalence of type IIA and heterotic vacua
implies in particular that their respective  moduli spaces
are identical  and  that a weak coupling computation
in a type II setting gives non-perturbative information about
the dual heterotic vacuum and vice versa.

In order to make contact with the heterotic prepotential of eq.~\Fhetet\
we need to compute the same quantity in type IIA vacua.
In the large volume limit of a Calabi--Yau manifold one has generically
$$
\Ftwo =  \coeff{1}{6} d_{\a\b\c} t_\a t_\b t_\c +
      {\rm worldsheet\ instantons} \ ,
\eqn\Ftwogen
$$
where $t_\a, \a=1,\ldots, \hone$ are the complexified K\"ahler moduli
(\ie\ $J=\sum_\a t_\a J_\a,$ $\Re t_\a >0$);
$d_{\a\b\c}\equiv \int J_\a\wedge J_\b \wedge J_\c$ are the classical
intersection numbers,
$J_\a\in H^{1,1}(Y,\ZZ)$ are the generators of the K\"ahler cone.

For such a vacuum to be dual to a perturbative heterotic vacuum
one of the $(1,1)$ moduli, say $t_s$,
has to be identified with the heterotic dilaton $S$.
In order for the  two prepotentials \Fhetet\ and \Ftwogen\ to coincide
the intersection numbers have to obey
$d_{sss} = d_{ss\a} =0$.
In addition, the higher derivative coupling $F_1$
obeys in the large volume limit of the type II vacua \BCOV
$$
F_1 = \sum_\a\,  c_2(J_\a)\  t_\a +
      {\rm worldsheet\ instantons} \ ,
\eqn\FoneCY
$$
where  $c_2(J_\a) \equiv \int c_2\wedge J_\a$
($c_2$ is the second Chern--class of the Calabi--Yau manifold).
Agreement with the heterotic $F_1$ of equation \Fonelarge\
implies $c_2(J_s) = 24\  (= \chi(K3))$.
These conditions (together with the `nefness' of the associated divisor)
imply that a  type IIA vacuum which is dual to a perturbative
heterotic vacuum necessarily has to be a K3-fibration \refs{\KLM,\AL}.
That is, there is a holomorphic map $Y\to\IP_1$ where the generic fiber
is a smooth $K3$.
However, not every K3-fibration has to be the dual of a perturbative
heterotic vacuum. It always has a  candidate modulus (namely $t_s$)
for the heterotic dilaton but some of the moduli might not couple
to this dilaton in the same way as the
perturbative heterotic moduli $C^i$ in eq.~\Fhetet.
This occurs precisely when the fiber degenerates and there exist
$(1,1)$ forms associated with the resolution of such degenerations \AL.
These moduli have to be identified as the type II dual of
the non-perturbative gauge fields $C'$ or additional
vector-tensor multiplets $\B$ introduced  in section~2.2.
It is important to keep in mind  that the one perturbative vector-tensor
multiplet which contains the dilaton
as well as the possible non-perturbative vector-tensor multipets
are mapped to honest vector multiplets in the dual type IIA vacua.

The previous discussion can be supplemented with  the additional
condition that the heterotic vacuum is toroidally compactified from $d=6$.
In this case the dual Calabi--Yau threefold has to be  an
elliptic fibration which is believed
to be the exact same Calabi--Yau threefold on which
F--theory is compactified and which captures the non-perturbative
physics of the six-dimensional heterotic vacua \refs{\vafa,\MV}.
In terms of the intersection numbers elliptic fibrations
satisfy $d_{ttt}=0, d_{tt\a} \neq 0$ \MV\ for some $\a$ where
we denote by $t_t$ the $(1,1)$ modulus of the elliptic curve.
In eq.~\Presult\ we learned that indeed the cubic polynomial
$P_3^\one$ obeys this condition if one identifies $t_t$ with the
radial modulus  of the torus $T$.\foot{The perturbative heterotic
string is completely  symmetric under the exchange  $T\leftrightarrow U$.
However, the identification of $T$ with the radius in eq.~\Trel\
chooses the asymptotic conditions on $T$ and $U$ and selects $\Re T > \Re U$.
Furthermore, the condition $d_{tt\a} \neq 0$ cannot be observed on the heterotic
side, since such couplings are ambiguous.}
Furthermore, if the six-dimensional heterotic vacuum has
additional tensor multiplets the $F_1$ (in $d=4$)
obeys eq.~\twelve\ and agreement with  \FoneCY\
implies $c_2(J_v) = -12$.

If the toroidally compactified 
heterotic vacuum  has a dilaton (and thus a weak coupling limit),
the elliptic fibration should  also be  a K3-fibration.
On the other hand, non-perturbative heterotic vacua with a
dilaton  frozen in the strong coupling region are dual to elliptic
Calabi--Yau threefolds which do not admit a  K3-fibration.
Finally, for the special case of heterotic vacua with equal instanton
numbers the discussion at the end of the previous section suggests that
the Calabi--Yau threefold should admit
two inequivalent  K3-fibrations corresponding to choosing
$S$ or $T$ as the heterotic dilaton or in other words choosing
a heterotic vacuum or its dual \refs{\AG,\MV}.
We now turn to a more detailed description of a few
explicit examples which display these properties.

%
\section{Construction of Calabi--Yau manifolds using toric geometry}
The vacua  we discuss explicitly  all have a description within
toric geometry (see e.g.~\nrf{\toric\AGM\HKT}\refs{\toric-\HKT}).
Specifically, we are looking at elliptic fibrations where the base is either
$\IP_2$, a Hirzebruch surface $\IF_n$ or blow-ups (of toric fixed points)
thereof, but we
restrict ourselves to the simplest cases, namely  $\IF_{0,1,2}$ as a base
with at most two blow-ups. We first give the toric description of the
base and then of the elliptically fibered Calabi-Yau manifold with this
base.

We characterize a toric surface in terms of a complete regular two-dimensional
fan. For $\IF_n$ the fan is generated by $\n_1=e_2,\,\n_2=e_1,\,
\n_3=-e_2,\,\n_4=-e_1+n e_2$ where $e_1,e_2$ are two-dimensional Euclidian
unit vectors. Other, combinatorically equivalent ways of drawing the fan
will be employed in some of the figures. 
Note the two independent relations $\n_1+\n_3=0,\,
\n_2+n \n_3+\n_4=0$. There are two so-called primitive collections
(see Batyrev in \toric):
${\cal P}_1=\{\n_1,\n_3\},\,{\cal P}_2=\{\n_2,\n_4\}$.
We can thus write $\IF_n$ as
$\IC^4-\{\{z_1=z_3=0\},\{z_2=z_4=0\}\}/(\IC^*)^2$
where $(\IC^*)^2$ acts as
$(z_1,z_2,z_3,z_4)\to(\lambda z_1,\mu z_2,\lambda\mu^n z_3,\mu z_4)$.
$\IP_2$ is described by the fan $\n_1=e_2,\,\n_2=e_1,\,\n_3=-(e_1+e_2)$
with the relation $\n_1+\n_2+\n_3=0$ and the primitive collection
${\cal P}=\{\n_1,\n_2,\n_3\}$. We thus write $\IP_2$ as the quotient
$\IC^4-\{z_1=z_2=z_3=0\}/\IC^*$ where $\IC^*$ acts as
$(z_1,z_2,z_3)\to(\lambda z_1,\lambda z_2,\lambda z_3)$. The fan for a blow up
is obtained by adding the generator $\n_i+\n_{i+1}$. To each generator we
can associate a divisor $D_i\simeq\IP_1$. They have intersection numbers
$D_i\cdot D_j=1$ for $|i-j|=1$, self-intersection number
$D_i\cdot D_i=a_i$ where $a_i$ is defined through the relation
$\n_{i-1}+\n_{i+1}+a_i \n_i=0$ ($\n_{N+1}\equiv \n_1$,
$N$ is the number of generators)  and zero intersection otherwise.
It is easy to see that a blow-up induces
the change
$(a_1,\dots, a_i,a_{i+1},\dots, a_N)\to(a_1,\dots,a_i-1,-1,a_{i+1}-1,\dots,a_N)$.
Conversely, we can blow-down the $\IP_1$'s with self-intersection
number $-1$ and still end up with a non-singular surface.
In this way we get $\IP_2$ from $\IF_1$. We can also easily
describe the transition $\IF_n\to\IF_{n\pm1}$ in terms of the
self-intersection numbers:
$(-n,0,n,0)\to(-n-1,-1,-1,n,0)\to(-(n+1),0,n+1,0)$ for
$\IF_n\to\IF_{n+1}$ and
$(-n,0,n,0)\to(-n,0,n-1,-1,-1)\to(-(n-1),0,n-1,0)$ for
$\IF_n\to\IF_{n-1}$. Here the first step is a blow up and the
second a blow down.
The toric diagrams for the transitions $\IF_1\leftrightarrow\IF_2$
are shown in fig.~1, with the self-intersection
numbers of the $\IP_1$'s included.
Since we can get $\IF_1$ from $\IP_2$ via blow-up and
$h^{1,1}(\IP_2)=1$ and every $\IP_1$ adds one (1,1)-form, we have
$h^{1,1}(B)=N-2$ where $B$ is the toric surface whose fan has $N$ generators.
If $B$ is the base of an elliptic Calabi-Yau manifold,
the number of tensor multiplets is $N-3$, according to eq.\Fcount. 
\vskip.3cm
$$
\eqalign{
&\epsfbox{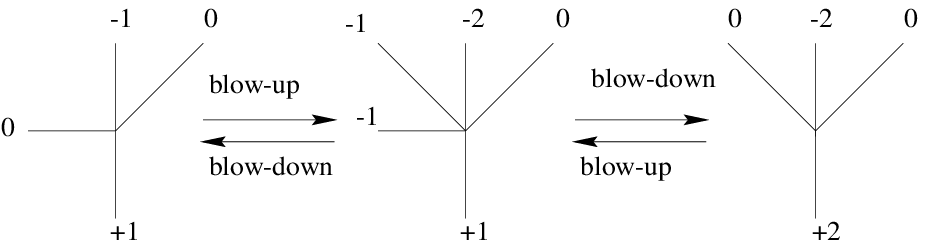}\cr
&Fig.~1: {\it The\ transitions}\ \IF_1\leftrightarrow\IF_2}
$$
\vskip.5cm
For the general (compact smooth) toric surface we can give a description
analogous to the one we have given above for $\IF_n$ and $\IP_2$. We can
write it as the quotient $(\IC^N-{\cal M})/(\IC^*)^{N-2}$ where the set
${\cal M}=\cup_{|i-j|\geq2}\{z_i=z_j=0\}$ is defined by the ${1\over2}N(N-3)$
primitive collections and the $(\IC^*)^{N-2}$ action is
$(z_i,z_{i+1},z_{i+2})\to(\lambda_i z_i,\lambda_i^{a_{i+1}}z_{i+1},
\lambda_i z_{i+2})$ for $i=1,\dots,N-2$.

For the construction of Calabi-Yau manifolds we use Batyrev's method 
of four-dimensional reflexive  polyhedra \Batyrev. Elliptic
fibrations are obtained by choosing polyhedra such 
that they contain a two-dimensional face
that can be triangulated to obtain the fan of one of the toric surfaces 
discussed above. 
In addition we also need to incorporate the combinatorial structure
dictated by the elliptic fiber. 

The models we treat in detail are
summarized, together with some related models, in the table.
The notation is as follows. We specify the base, which is a Hirzebruch surface 
with up to two blow ups. Each blow-up corresponds
to an additional tensor multiplet on the heterotic side. 
It  results from an
$E_8$ instanton shrunk to zero size which can occur in either one of the two
$E_8$ factors (indicated by a subscript);
this lowers the instanton number of the corresponding
factor by one unit. We can reach a situation with instanton numbers
$(n_1,n_2)$ either by starting with $(n_1+1,n_2)$ or $(n_1,n_2+1)$ and
shrinking an instanton in the first and the second factor, respectively.
We thus list only those blow-ups of $\IF_n$ which are not also blow-ups
of $\IF_{n-1}$.
The required Hodge numbers of the type II vacuum are obtained via eqs.~\dimtt. 
The polyhedra specified in the last column 
are either from or extensions of those of ref. \CF.
To describe the base we introduce the vertices
$$
\nu_1=(0,1,2,3),\,
\nu_2=(1,1,2,3),\,
\nu_3=(1,0,2,3),\,
\nu_4=(1,-1,2,3),
$$
$$
\nu_5=(0,-1,2,3),\,
\nu_6=(-1,-1,2,3),\,
\nu_7=(-1,0,2,3),\,
\nu_8=(-1,1,2,3),
$$
$$
\nu_9=(1,2,2,3),\,\nu_{10}=(0,0,2,3)\,.
$$
The first parenthesis in the last column of the table 
specifies the base by listing its vertices.
In addition to those listed there is always the vertex $\nu_{10}$.
In general there are several polyhedra leading to Calabi-Yau manifolds with
the same Hodge numbers and the same combinatorical structure concerning
the base. For instance, for the $\IF_2$ models we can either choose
$(\nu_1,\nu_5,\nu_7,\nu_9)$ or $(\nu_1,\nu_2,\nu_5,\nu_8)$ to specify the base.
Opening up $SU(2)$'s requires modification of the polyhedron by adding
extra vertices. They are among
$$
\rho_1=(0,-1,1,2),\,\rho_2=(0,1,1,2),\,
\rho_3=(0,-1,0,1),\,\rho_4=(0,1,0,1)\,,
$$
and specified as the entries of the second parenthesis.
In addition, all polyhedra contain the vertices
$$
\mu_1=(0,0,-1,0),\quad \mu_2 = (0,0,0,-1)\,,
$$
and the origin $(0,0,0,0)$. We have not specified vertices on
faces of codimension one.
Polyhedra for higher rank gauge groups can be found in
\CF\ and \BKKMSV.

The convex hull (denoted below by conv) of the vertices
$(\mu_1,\mu_2,\nu_{10})$
is the two-dimensional
polyhedron corresponding to the torus which is a degree $6$
hypersurface in $\IP(1,2,3)$. This is the generic elliptic fiber of
the models considered. If we add the vertices $(\nu_1,\nu_5)$
or, alternatively, $(\nu_3,\nu_7)$ we get the
three-dimensional polyhedron for the
degree 12 hypersurface in $\IP(1,1,4,6)$, which is a $K3$. If we add
$\rho_1$ (or $\rho_2$) we have a $K3$ fibration in two different ways.
There is still the $K3$ associated to the polyhedron
${\rm conv}(\mu_1,\mu_2,\nu_3,\nu_7)$, but the second $K3$ is now
given by the polyhedron ${\rm conv}(\mu_1,\mu_2,\nu_1,\nu_5,\rho_1)$.

For a given polyhedron,
the Calabi-Yau manifold, or, more precisely, the
toric variety in which it is a hypersurface, is specified by  a  particular
triangulation of the polyhedron. Here we consider only
 regular triangulations
which take into account all the vertices except those on faces of
codimension one and where all simplices contain the origin.
Such triangulations correspond to  Calabi-Yau phases
of the underlying conformal field theory.
There are in general several possible Calabi--Yau phases which
generically lead to topologically different Calabi-Yau manifolds \AGM.
Their  Hodge numbers are the same, but the intersection numbers
and the instanton
numbers are different. Below we  only specify the triangulation of the
two-dimensional face in the $(x_3,x_4)=(2,3)$ plane.
The question when different triangulations lead to the same Calabi-Yau
hypersurface has been addressed in \BKKMI. The different triangulations
of a given polyhedron that we consider always lead to distinct models.

Using the methods outlined in \PF\
we compute  $c_2(J_\a)$ and the prepotential
for some of the models specified in the table.\foot{We would
like to thank S.~Katz and A.~Klemm for providing computer codes implementing
parts of the computations.}
{}From our previous discussion we know
that those $J_\a$ with $c_2(J_\a)=24$
are candidates for the dual of the heterotic dilaton.
In addition, using eqs.~\Fhetet--\twelve\
we can also identify the six-dimensional heterotic origin of the
four-dimensional
vector multiplets: whether they arise from tensor multiplets, perturbative
or non-perturbative vector multiplets.

\vfill\eject
\def\tensor{{\rm tensor}}
\overfullrule=0pt
\vskip.3cm
{\footfont
\centerline{
\vbox{\offinterlineskip
\halign{&$\quad#\quad$&
\vrule\strut$\quad#\quad$&
\vrule\strut$\quad#\,\,$&
\vrule width 0.8pt\strut$\quad#\,\,$&
\vrule\strut$\quad#\quad$&
\vrule\strut$\quad#\quad$&
\vrule\strut$\quad#\quad$&
\vrule\strut$\quad#\quad$\cr
\#&{\rm model}&(n_1,n_2)  &h^{1,1}&h^{2,1}&-\chi &\Delta^*\cr
\noalign{\hrule}
&&&&&&\cr
1&{\bf P_2} && 2  & 272 &   544 & (2,5,7)  \cr
&&&&&&\cr
\noalign{\hrule}
&&&&&&\cr
2&\IF_0 & (12,12)& 3 & 243  & 480 &(1,3,5,7)\cr
3&\IF_1 & (11,13)& 3 & 243  & 480 &(1,2,5,7)\cr
4&\IF_2 & (10,14)& 3 & 243  & 480 &(1,2,5,8)\cr
&&&&&&\cr
\noalign{\hrule}
&&&&&&\cr
5&{\IF_0+SU(2)_1} & (12,12)& 4 & 214  & 420 & (1,3,5,7)(1)\cr
6&{\IF_0+\tensor_1} & (11,12)& 4 & 214  & 420 &(1,2,3,5,7)\cr
7&{\IF_1+SU(2)_1} & (11,13)& 4 & 226  & 444 & (1,2,5,7)(2)\cr
8&{\IF_1+SU(2)_2} & (11,13)& 4 & 202  & 396 & (1,2,5,7)(1)\cr
9&{\IF_1+\tensor_1} & (10,13)& 4 & 214  & 420 &(1,2,5,7,8)\cr
10&{\IF_2+SU(2)_1} & (10,14) & 4 & 238 & 468 & (1,2,5,8)(2)\cr
11&{\IF_2+SU(2)_2}& (10,14)& 4 & 190 & 372 & (1,2,5,8)(1) \cr
12&{\IF_2+\tensor_1}& (9,14)& 6 & 222  & 432 &{(1,2,5,8,9)}\cr
&&&&&&\cr
\noalign{\hrule}
&&&&&&\cr
13&{\IF_0+SU(2)_1\times SU(2)_2} & (12,12)& 5 & 185 & 360
&(1,3,5,7)(1,2)\cr
14&{\IF_0+SU(2)_1\times SU(2)_1} & (12,12)& 5 & 185 & 360
&(1,3,5,7)(1,3)\cr
15&{\IF_0+SU(2)_1+\tensor_1} & (11,12)& 5 & 197 & 384
&(1,2,3,5,7)(2)\cr
16&{\IF_0+SU(2)_1+\tensor_2} & (12,11)& 5 & 185  & 360
&(1,2,3,5,7)(1)\cr
17&{\IF_0+2\ {\rm tensors}} & {(10,12)\atop(11,11)}& 5 & 185  & 360
&(1,2,3,5,6,7)\cr
&&&&&&\cr
\noalign{\hrule}
&&&&&&\cr
18&{\IF_1+SU(2)_1\times SU(2)_2}& (11,13)& 5 & 185 & 360&
(1,2,5,7)(1,2) \cr
19&{\IF_1+SU(2)_2\times SU(2)_2}& (11,13)& 5 & 165  & 320
&(1,2,5,7)(1,3)\cr
20&{\IF_1+SU(2)_1+\tensor_1} & (10,13) & 5 & 209 & 408 &  (1,2,5,7,8)(2)\cr
21&{\IF_1+SU(2)_2+\tensor_1} & (10,13)& 5 & 173  & 336
&(1,2,5,7,8)(1)\cr
22&{\IF_1+\tensor_1+\tensor_1} & (9,13)& 7 & 193 & 372  &{(1,2,5,7,8,9)} \cr
&&&&&&\cr
\noalign{\hrule}
&&&&&&\cr
23&{\IF_2+SU(2)_2\times SU(2)_2} & (10,14) & 5 & 145 & 280 & (1,2,5,8)(1,3)\cr
24&{\IF_2+SU(2)_1\times SU(2)_2} & (10,14)& 5 & 185 & 360&
(1,2,5,8)(1,2) \cr
25&{\IF_2+SU(2)_2+\tensor_1} & (9,14)& 7 & 169 & 324
&{(1,2,3,8,9)(2)} \cr
26&{\IF_2+\tensor_1+\tensor_1} & (8,14)& 9 & 213 & 408&{(1,2,5,8,9)(1)} \cr
}}}}
\vfill\eject

%
\section{Vacua with $N_V=3$}
Let us first concentrate on perturbative
heterotic vacua where the entire gauge symmetry is
Higgsed away.  As discussed in section~2.1
this is possible for instanton numbers
$n >9$ and using \inumber\ reveals   the three possibilities
$(n_1,n_2)=(12,12), (11,13), (10,14)$. 
Each of these instanton numbers specifies a heterotic vacuum with
spectrum $(\nh,\nv,\nt)=(244,0,1)$ in six dimensions
and $(\nh,\nv,\nt)=(244,2,1)$ in the toroidally compactified $d=4$  vacuum.
Using \dimtt\ and the fact that a heterotic vector-tensor multiplet  is
mapped to
a vector multiplet in the dual type II vacuum we learn that the
Calabi--Yau threefold needs to have
$(h^{1,1},h^{2,1})=(3,243)$.
Calabi--Yau compactifications with these Hodge numbers
have been discussed previously in refs.~\refs{\KLM, \MV, \AG}.
They are elliptic with bases
$\IF_0,\,\IF_1$ and $\IF_2$, respectively \MV.
\vskip.3cm
$$
\eqalign{
&\epsfbox{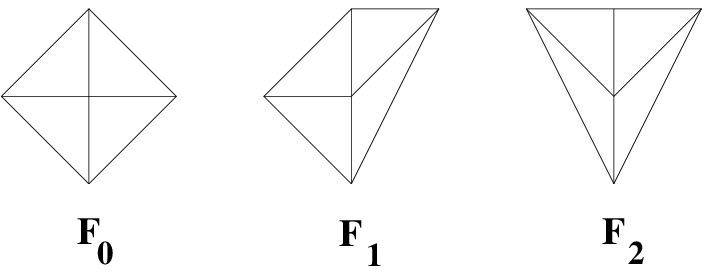}\cr
&{\rm Fig. 2}: {\it The\ toric\ diagrams\ for\ the\ surfaces}\
\IF_0,\,\IF_1,\,\IF_2.
}
$$

Choosing ${\IF_0}$ as a base  (model 2 in the table)
we find $c_2(J_\a)=\{92,24,24\}$ which is a `double'
K3-fibration as one has two choices for the base of the $K_3$ fibration
(or, equivalently, there are two
candidates for the dilaton). The fact that this  threefold  is  a
double fibration can also easily be seen
from its toric description in that there are two
ways to embed the polyhedron corresponding to the $K3$
in the polyhedron specified in the table
(see also ref.~\AG).
For the classical prepotential we find
$$
\Ftwo =  \coeff 43\,t_{1}^{3} + t_{{1}}^{2}t_{{2}}+t_{{1}}^{2}t_{{3}}
       +t_{{1}}t_{{2}}t_{{3}}\,,
\eqn\Fzero
$$
which   is completely symmetric under the exchange of
$t_2$ and $t_3$;  this  corresponds to an exchange of the two $\IP_1$'s
which serve as the base of the two alternative $K3$ fibrations.
This symmetry can also be checked for the entire prepotential including
the instanton corrections.
Therefore, this vacuum should be identified as the type II dual of the
heterotic $(n_1,n_2) = (12,12)$ vacuum which is expected to have this  symmetry
as a consequence of the heterotic-heterotic duality.
The identification between the type II and heterotic moduli
$$
t_1 = U, \qquad  t_2 = T - U, \qquad t_3 = S - U,
\eqn\Fzerosubs
$$
inserted into \Fzero\ reveals
$$
\Ftwo\ =\ STU + \coeff13 U^{3}  \ .
\eqn\Fzerof
$$
This prepotential is consistent with the heterotic $\Fhet$ defined in
\Fhetet--\Presult\
since the condition $t_2 >0$ chooses
$\Re T > \Re U$ and renders \Fzerof\ and \Presult\ consistent \CCLM.
Also, we need $\Re S>\Re U$, which is indeed the condition for being in
the perturbative regime.
Obviously one could have exchanged $S$ and $T$ in \Fzerosubs\
without altering $\Ftwo$  in \Fzerof\ in accord  with the expected $S-T$
exchange symmetry.

This symmetry was first observed in \KLM\ for the degree 24
hypersurfaces in $\IP(1,1,2,8,12)$ which, in our notation,
is the same as vacuum 4 which has   $\IF_2$ as base.
One finds
$c_2(J_\a)=\{92,48,24\}$ and
$$
\Ftwo =\coeff43 \, t_{{1}}^{3} + 2\,t_{{1}}^{2}t_{{2}}+t_{{1}}t_{{2}}^{2}
+t_{{1}}^{2}t_{{3}}+t_{{1}}t_{{2}}t_{{3}}\ .
\eqn\Ftwoo
$$
With the substitution $t_3 \to t_3-t_2$ this turns into the
prepotential of the
$\IF_0$ model and furthermore the equivalence continues to hold when the
instanton corrections are included and the full prepotentials are
compared.\foot{The 
$c_2(J_\a)$ also match since a change $t_\a\to A_{\a\b}t_\b$
induces  $J_\a\to A^{-1}_{\b\a}J_\b$. }
The relation between the K\"ahler moduli of these two models means that
the K\"ahler cone of the $\IF_2$ model is a subcone of the K\"ahler cone
of the $\IF_0$ model.
The heterotic dual of the $\IF_2$ model has been identified as
the vaccum with instanton numbers  $(n_1,n_2) = (10,14)$ 
which is in the same moduli space as the $(12,12)$ vacuum \refs{\MV,\AG}.

Choosing
$\IF_1$ as the base (model 3)  we compute $c_2(J_\a)=\{92,36,24\}$ and the
classical prepotential
$$
\Ftwo =
\coeff43 \,t_{{1}}^{3}+\coeff32 \,t_{{1}}^{2}t_{{2}}
    +\coeff12 t_{{1}}t_{{2}}^{2}+t_{{1}}^{2}t_{{3}}+t_{{1}}t_{{2}}t_{{3}}\ .
\eqn\Foneone
$$
In this case there is also a  linear transformation
of the moduli which transforms \Foneone\
into \Fzero\ but the coefficients of the transformation are not all integer:
$(t_1,t_2,t_3)\to (t_1,t_2, t_3+\coeff12 t_2)$.
Inspection of the instanton contributions to the prepotential shows that
the expansion in $q_i=e^{-2\pi t_i}$ would not be in {\sl integer}
powers of $q_2$.
This vacuum  is physically different from  the  $\IF_0$ and $\IF_2$ vacua
and the instanton corrections do not agree; it
has been identified with the heterotic 
$(n_1,n_2) = (11,13)$ vacuum. 
The substitution\foot{In 
identifying the heterotic variables non-integer transformations
are generically allowed. In particular the dilaton is ambiguous as we
discussed below eq.~\Fhetint. However, the fields  that couple
to the dilaton ($T,U,C$) may only be shifted such as to 
respect the correspondence with eq.~\Fhetet. 
Similarly, eq.~\Fonelarge\
constrains the dilaton dependent shifts of all variables.}
$$
t_1 = U, \qquad t_2 = T - U,  \qquad   
t_3 = S -\coeff12 T - \coeff12 U,
\eqn\Fonesubs
$$
into \Foneone\ gives
$$
\Ftwo= STU + \coeff13  U^{3} \ ,
\eqn\Foneonef
$$
consitent with \Fhetet. In all  three vacua based on $\IF_{0,1,2}$ 
the heterotic weak coupling $S\to \infty$
limit corresponds to the $t_3 \to\infty$ limit in the type II vacuum
in which the instanton corrections  are identical.
This says that perturbative heterotic
prepotentials of the three models coincide.
Conversely, a purely perturbative check of  dual vacua
as has been performed for example in refs.~\refs{\KLM,\KLT-\CCLM}
is unable to distinguish between these models. Additional
non-perturbative input -- namely the embedding of the instantons
and the resulting strong coupling behaviour -- is required
to uniquely identify the dual pairs.\foot{The 
same phenomenon has been observed by
Berglund, Katz, Klemm and Mayr \BKKM\ and we are grateful for 
communication of these results prior to publication.}

The polyhedron of the $\IF_1$ model also admits a second
triangulation which is obtained via a flop in the two-dimensional
face describing the base; see also the discussion in \MV. (The flop is 
shown in fig.~3, which we discuss in the next section.)
The resulting model has $c_2(J_\a)=\{92,102,36\}$ which shows that
it is not a $K3$ fibration as can also be seen from the toric diagram.
Its classical prepotential is
$$
\Ftwo =\coeff43 t_{{1}}^{3} + \coeff32  t_{{2}}^{3}
+\coeff92 t_{{1}}^{2} t_{{2}} + \coeff92 t_{{1}} t_{{2}}^{2}
+ \coeff32 t_{{1}}^{2} t_{{3}} + \coeff32 t_{{2}}^{2} t_{{3}}
+ \coeff12  t_{{1}} t_{{3}}^{2} + \coeff12 t_{{2}} t_{{3}}^{2}
+3 t_{{1}} t_{{2}} t_{{3}}\ .
\eqn\Fonetwo
$$
If  we set $t_1=0$ we obtain  the
prepotential of the two-parameter model (model 1)
with $\IP_2$ as the base.
The transition from model 3 to model 1 involves
shrinking a four cycle  which can only be done  after performing the flop.
In the flopped vacuum 
one can find a basis where one variable completely decouples.
This corresponds to a divisor which does not intersect any other divisor
in this new basis. This divisor will then be shrunk. Indeed, substituting
$$
t_1 = \BX\ , \quad t_2 = U- \BX\ ,\quad t_3 = T-\coeff32 U\ ,
\eqn\Fonesubss
$$
gives
$$
\Ftwo=
\coeff38 U^3 +\coeff12 UT^{2} - \coeff16 \BX^3\ .
\eqn\Fonetwof
$$
\vfill\eject

\section{Vacua with $\nv=4$}
By adding one additional vertex to the polyhedra of the three-parameter 
models in such a way that the resulting polyhedra stay reflexive 
one constructs vacua with $\nv=4$. This can be done in 
different ways leading to models 5--11 in the table.
By blowing up the base the additional vector multiplet is
the type II dual of a vector-tensor multiplet as is expected from the discussion
in section~2.2.
Alternatively,  adding  a vertex without
touching the base results in an  additional $U(1)$ vector multiplet which
parameterizes the Coulomb branch of an $SU(2)$ gauge symmetry.
\vskip.3cm
$$
\eqalign{
&\qquad\epsfbox{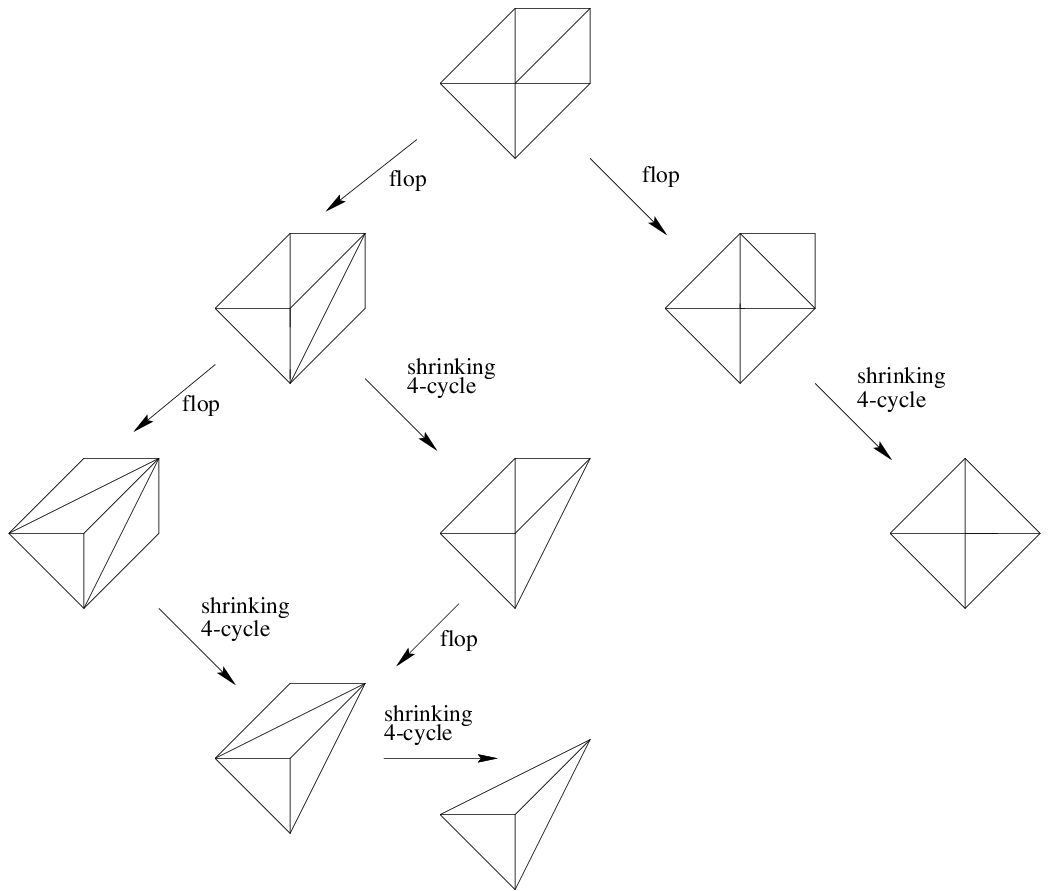}\cr
&{\rm Fig. 3}: {\it Base\ of\ Vacuum\ 6\ and\ the\ relations\ with\
vacua\ 1,\ 2\ and\ 3}
}
$$
We can either blow up $\IF_0$ or $\IF_1$ to arrive at the base of vacuum 6.
The self-intersection numbers of the $\IP_1$'s are
$(-1,-1,-1,0,0)$. The toric diagram of the base together
with its triangulation is depicted at the top of fig.~3 and
we immediately see that again there will be
two candidates for the dilaton.
We find
$c_2(J_\alpha)=\{36,24,24,82\}$ and  the prepotential
$$
\Ftwo =\coeff76 t_{{4}}^{3} + t_{{4}}^{2}t_{{2}}
+ \coeff3 2 t_{{4}}^{2}t_{{1}}
+ \coeff12  t_{{4}}t_{{1}}^{2}
+ t_{{4}}^{2}t_{{3}}+t_{{4}}t_{{2}}t_{{1}}
+ t_{{4}}t_{{2}}t_{{3}}+t_{{4}}t_{{1}}t_{{3}}\ .
\eqn\Fzerone
$$
The expected symmetry $t_2\leftrightarrow t_3$ is again manifest in $\Ftwo$ but
also extends to the entire prepotential including the worldsheet instantons.
To make contact with the heterotic prepotential we substitute
$$
t_1=\Bb-\coeff12 U,\quad
t_2 = T-\Bb-\coeff12 U, \qquad
t_3=S-\Bb-\coeff12 U,  \qquad
t_4 = U,
\eqn\Fzeronesubs
$$
into \Fzerone\ and obtain
$$
\Ftwo=STU -\coeff12  U\Bb^2 +  \coeff7{24}  U^{3} \ .
\eqn\Fzeronef
$$
Again this is consistent with the dual heterotic vacuum. $\Bb$ does not couple
to the dilaton and thus cannot be a vector multiplet of a
perturbative heterotic vacuum. Its couplings to $T$ and $U$ are consistent with
eq.~\purecube\ and furthermore, the change of variables \Fzeronesubs\
changes the $c_2(J_\alpha)$ such that  $c_2(J_V)=-12$ consistent
with \twelve.  Thus, we identify $\Bb$ as the type II dual of a  heterotic
vector-tensor multiplet.
Let us also note that the coefficient of the $U^3$ term has changed
compared to the three parameter models and is no longer
in agreement with \Presult. However,
\Presult\ is valid  in perturbative heterotic vacua but
here we have an additional vector-tensor multiplet and are thus outside
the validity of the computation of ref.~\HM.
However, in all models we considered this coefficient is given by
$\coeff{-1}{24\cdot 60}\, \chi = \coeff{1}{24} (9- \nt)$ 
in the basis choosen in \Fzeronef\ and where 
$\nt$ counts the dilaton and the number of $\Bb$'s (the $V_X$'s do not
contribute to this coefficient). 
It would be interesting to confirm this result by an
independent computation on the heterotic side.

The transition from vacuum 6 to vacua 1, 2 or 3
procedes through an intermediate Calabi--Yau phase
which  involves a flop on the polyhedron of model 6.
There are two inequivalent  such flops which are indicated in the second row
of fig.~3. In the `flopped phase'
a four cycle can be shrunk and one reaches model 2 or 3 respectively.
The triangulation on the left side admits a second flop and after
shrinking two four cycles one arrives at vacuum~1 which we already
discussed briefly in the previous section.

In terms of the prepotential  one observes that neither \Fzero\ nor \Foneone\
can be obtained from \Fzerone\ by  simply setting one  of the parameters
to zero.
However, in
the flopped phase for example on the right  hand side in  fig.~3 one finds
$c_2(J_\alpha)=\{92,24,24,82\}$
(indicating that there are still two dilatons) and
$$
\eqalign{
\Ftwo =&\ \coeff43 t_{{1}}^{3} + \coeff76 t_{{4}}^{3}
+ t_{{3}}t_{{1}}^{2} + t_{{2}}t_{{1}}^{2} + t_{{3}}t_{{4}}^{2}
+ 4\,t_{{1}}^{2}t_{{4}}
+ t_{{2}}t_{{4}}^{2}+4\,t_{{1}}t_{{4}}^{2}+t_{{3}}t_{{2}}t_{{1}}\cr
+&\  t_{{3}}t_{{2}}t_{{4}}+2\,t_{{3}}t_{{1}}t_{{4}}+2\,t_{{2}}t_{{1}}t_{{4}}\ .
}
\eqn\Fzeroflop
$$
Now setting   $t_4=0$ results in the prepotential \Fzero.
Furthermore, after the substituting
$(t_1,t_2,t_3,t_4)\to(-t_1,t_2+t_1,t_3+t_1, t_4+t_1)$ into \Fzeroflop\
the two prepotentials \Fzerone\ and  \Fzeroflop\ only differ by a term $\coeff16
t_1^3$
which  is exactly what one expects after a flop \EWc.
The transformation of the parameters is obtained by considering the
relation between the generators of the Mori cones of the two triangulations
leading to the two models.
In the flopped  phase the heterotic variables are
$$
t_1=U-\BX, \qquad
t_2=T-U, \qquad
t_3=S-U, \qquad
t_4=\BX,
\eqn\Fzeroflopsubs
$$
which when substituted into \Fzeroflop\ results in
$$
\Ftwo = SUT  +\coeff13  U^{3} - \coeff16  \BX^{3}  \ .
\eqn\Fzeroflopf
$$
(Again we see that by putting $\BX=0$ one obtains \Fzerof.)

In the heterotic vacuum the transition between vacuum 6
and vacuum 2 or 3 corresponds to leaving the non-perturbative
Coulomb branch with the additional tensor multiplet and returning to the
perturbative vacua with instanton numbers $(12,12)$ or $(11,13)$ and only
one tensor multiplet.
The physical interpretation of the flopped phase in the heterotic vacua
is less straightforward. 
In six space-time dimensions this phase is not part of the 
F-theory moduli space and thus does not correspond to a 
heterotic vacuum in $d=6$ \MV.
In five dimensions there is a phase transition
associated with a flop; a hypermultiplet becomes massless
and induces a change in the Chern--Simons interactions of the gauge fields
which results in a  shift in the prepotential \EWb.
Comparing the prepotentials \Fzeronef\ and \Fzeroflopf\ we indeed see that
the Chern--Simons interactions of the vector-tensor multiplet has changed.
In  \Fzeronef\ $\Bb$ only appears quadratic in agreement with the dimensional
reduction from six dimensions \FMS. However, in \Fzeroflopf\
the vector-tensor completely decouples and has no couplings to any
of the other vector fields. This is precisely the prepotential obtained
in four dimensions
in ref.~\faux\ where the tensor fields of the vector-tensor multiplet only
couples to its own vector (and the graviphoton).
This behaviour -- the decoupling of the vector-tensor multiplet --
we observed in all flopped phases of Calabi--Yau threefolds
with blown up $\IF_{0,1,2}$ as a base (appendix A).
Furthermore, in all cases we find $c_2(J_{\BX}) = -10$ and the coefficient
in front of the $U^3$ term also changes by $1/24$.
In $d=4$ the flopped phases definitely are part of the moduli space
but it would be nice to understand their physics on the
heterotic side in more detail.

Let us discuss another blow up of $\IF_1$.
Recall that the base of vacuum 6 (top of fig.~3)
is a blow-up of $\IF_0$ but it can also
be viewed as a blow up of $\IF_1$.  There is a second
blow-up of $\IF_1$ which can also be viewed as a blow-up of
$\IF_2$ (fig.~4).
For this blow-up the self-intersection numbers of the
$\IP_1$'s are $(-2,0,1,-1,-1)$ and it is the base of vacuum 9.
\vskip.3cm
$$
\eqalign{
&\qquad\epsfbox{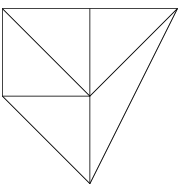}\cr
&{\rm Fig. 4}: {\it Base\ of\ vacuum\ 9}
}
$$ 
\vskip.3cm
\noindent 
This vacuum has
$c_2(J_\a)=\{36,48,24,82\}$ and
$$
\Ftwo=\coeff76\,t_{4}^{3} + \coeff32\,t_{1}t_{4}^{2}
+\coeff12 t_{1}^{2}t_{4} + 2\,t_{4}^{2}t_{2}+t_{4}t_{2}^{2}
+t_{4}^{2}t_{3}+2\,t_{1}t_{4}t_{2}+t_{1}t_{4}t_{3}
+t_{4}t_{2}t_{3}\, .
\eqn\Foneblow
$$
Substituting $(t_1,t_2,t_3,t_4)\to(t_1,t_2,t_3-t_2,t_4)$ shows the
equivalence with the prepotential of the blown up $\IF_0$ model \Fzerone;
it extends to the instanton contributions.
This is an immediate consequence of the equivalence of vacua 2 
and 4.\foot{Another possible blow-up of $\IF_2$ has self-intersection
numbers $(-3,-1,-1,2,0)$. This leads  to instanton numbers
$(n_1,n_1)=(9,14)$ and we can no longer completely break the first
$E_8$ factor. It turns out \DMW\ that we are left with an unbroken $SU(3)$.
This leads in the four-dimensional situation to alltogether
six vectors and a dual type II  model on a Calabi-Yau manifold with
Hodge numbers $(h^{1,1},h^{2,1})=(6,222)$. This is model 12 in the table.}

So far we have considered models with $\nv=4$ where the fourth
vector multiplet originates from a six-dimensional
tensor multiplet.
Let us now consider the Coulomb branch of vacua with an $SU(2)$ gauge
symmetry.
Vacuum 5 again  has  a double $K3$ fibration.
Thus  we expect two candidates for the dilaton
and an $S-T$ exchange symmetry inherited from the six-dimensional
heterotic-heterotic duality. However, since there is a gauge symmetry
we also expect to observe  the exchange of perturbative with non-perturbative
gauge fields. Indeed, there are now two {\sl different} $K3$ surfaces
due to the additional vertex $\rho_1$   as can be seen from the polyhedron
(in model 2 the $K3$'s were identical; c.f.~discussion in section 3.1).
We find $c_2(J_\a)=\{92, 24,24,248\}$ and the classical prepotential
$$
\eqalign{
\Ftwo =&\ \coeff{94}{3}  t_{4}^{3} + \coeff43 t_{1}^{3}
+8\,t_{3}t_{4}^{2}+9\,t_{2}t_{4}^{2}+34\,t_{4}^{2}t_{1}+t_{3}t_{1}^{2}
+t_{2}t_{1}^{2}+12\,t_{4}t_{1}^{2} \cr
&\ \ +3\,t_{2}t_{3}t_{4}+t_{2}t_{3}t_{1}
+6\,t_{3}t_{4}t_{1}+6\,t_{2}t_{4}t_{1} \ .
}
\eqn\Fpert
$$
$t_2$ and $t_3$ are both candidates for the heterotic dilaton but
$\Ftwo$ is not symmetric with respect to their interchange.
Substituting
$$
t_1=U-3C,\qquad
t_2=T-U,\qquad
t_3=S-U\qquad
t_4=C,
\eqn\Fpertsubs
$$
gives
$$
\Ftwo= S(TU-C^2)  + \coeff43 C^3 - U C^2 +  \coeff13  U^3\ .
\eqn\Fpertf
$$
With $S$ chosen as the dilaton  $C$ couples like a perturbative 
$U(1)$ (cf.~\Fhetet).
Since  $c_2(J_T) =24$ also $T$ can serve as the dilaton but with
respect to $T$ the  multiplet $C$ couples like a non-perturbative gauge field.
This confirms the prediction of the heterotic-heterotic duality in that
$\Ftwo$ is symmetric under a $S\leftrightarrow T$ exchange if
at the same time perturbative and non-perturbative gauge fields are
interchanged. The last two terms in \Fpertf\ are consistent with 
\Presult\ since the coefficient of the $\b$-function 
$b=12$ for the number of doublets computed in \idimg.

Let us close this section with vacuum  8. Here we choose a triangulation of the
polyhedron such that the resulting Calabi-Yau is a $K3$ fibration.
This choice is not unique; we picked the one with
$c_2(J_\a)=\{92,36,24,236\}$. For the prepotential we find
$$
\eqalign{
\Ftwo&=\coeff{88}3 t_4^3+8 t_4^2 t_3+\coeff{25}2 t_4^2 t_2+3 t_4 t_2 t_3+
\coeff32 t_4 t_2^2+33 t_4^2 t_1\cr
&+6 t_1 t_3 t_4+9 t_1 t_2 t_4+t_2 t_3 t_1+\coeff12 t_2^2 t_1
+12 t_4 t_1^2+t_3 t_1^2+\coeff32 t_2 t_1^2+\coeff43 t_1^3\,.}
\eqn\FFpert
$$
Via the substitution
$$
t_1=U-3 C,\qquad
t_2=T-U,\qquad
t_3=S-{1\over2}T-{1\over2}U,\qquad
t_4=C,
\eqn\FFpertsubs
$$
\FFpert\  turns into
$$
\Ftwo=S(TU-C^2) +{7\over3}C^3-{3\over2}U C^2-{1\over2}T C^2+{1\over3}U^3\,.
\eqn\FFpertf
$$
This vacuum has $b=18$ which once more establishes consistency 
with eq.~\Presult.
%
%
\section{Models with $N_V=5$}

We now consider models with five vector multiplets.
They can either arise from two, one or zero six-dimensional tensor multiplets.
We start with vacuum 17 which has as a base the $\IF_0$ surface blown
up twice and therefore we expect the dual heterotic vacuum to have
two tensor multiplets.  These can arise by shrinking two  instantons either
in the same or in different $E_8$ factors and thus the heterotic vaccum
has instanton numbers  $(11,11)$ or $(10,12)$.
There are three distinct double blow-ups of $\IF_0$, the difference is
visible from the self-intersection numbers of the $\IP_1$'s 
corresponding to the six generators.
It is straightforward  to construct corresponding four-dimensional
polyhedra, each with Hodge numbers $(\hone=5, \htwo=185)$.
We find that the full prepotential (including worldsheet instanton
corrections) of the three different blow-ups
of $\IF_0$ are equivalent.
This seems to imply that the two heterotic vacua are identical.
For the choice of base indicated  in the table we find
$c_2(J_\a)=\{72,24,36,24,24\}$ and
$$
\eqalign{
\Ftwo=&\ t_1^3+t_1^2 t_2+\coeff32 t_1^2 t_3+t_1 t_2 t_3+ \coeff12 t_1 t_3^2
+t_1^2 t_4+t_1 t_2 t_4\cr
&\ +t_1 t_3 t_4+t_1^2 t_5+t_1 t_2 t_5+t_1 t_3 t_5+t_1 t_4 t_5 \ , }
\eqn\doubleup
$$
which is completely symmetric in $t_2,t_4,t_5$.
This observation extends to the instanton contributions to the
prepotential.
Via the substitution
$$
t_1=U,\ \ \
t_2=S-\coeff12 T-\coeff12 U,\quad
t_3=T-V_Y-W_Y,\quad
t_4=V_Y-\coeff12 U,\quad
t_5=W_Y-\coeff12 U,
\eqn\doubleupsubs
$$
we arrive at
$$
\Ftwo=STU -\coeff12 U V_Y^2-\coeff12 U W_Y^2+\coeff14 U^3\ .
\eqn\doubleuphet
$$
$V_Y$ and $W_Y$ couple like vector-tensor multiplets and
in terms of the heterotic varibles we also find $c_2(J_V)=c_2(J_W)=-12$
consistent with \twelve. Note that the $U^3$ term is 
in accord with its coefficient being $\coeff1{24 }(9-\nt)$.

We next study model 16 which  has a $SU(2)$ and a  tensor
connected to a small instanton in the other $E_8$ factor.
We find $c_2(J_\a)=\{24,36,24,218,82\}$
and
$$
\eqalign{
\Ftwo&=3t_1t_2t_4+ \coeff32 t_2^2t_4+3t_1t_3t_4+3t_2t_3t_4+8t_1t_4^2
+ \coeff{25}{2} t_2t_4^2+9t_3t_4^2+ \coeff{161}{6} t_4^3\cr
&+t_1t_2t_5+ \coeff12 t_2^2t_5+t_1t_3t_5+t_2t_3t_5+6t_1t_4t_5
+9t_2t_4t_5+6t_3t_4t_5+ \coeff{59}{2} t_4^2t_5\cr
&+t_1t_5^2+\coeff{3}{2} t_2t_5^2+t_3t_5^2+\coeff{21}{2}t_4t_5^2
+\coeff{7}{6} t_5^3 \ .}
\eqn\aaa
$$
There are two candidates for the dilaton, but the classical prepotential
is {\sl not} symmetric in $t_1$ and $t_3$.
Similar to model~5 this had
to be expected from heterotic-heterotic duality since the two  dilatons
have to distinguish perturbative and non-perturbative gauge fields.
Subtituting
$$
t_1=S-\coeff12 U-V_Y,\quad
t_2=V_Y-\coeff12 U,\quad
t_3=T-\coeff12 U-V_Y,\quad
t_4=C,\quad
t_5=U-3C,
\eqn\aaasusone
$$
we get
$$
\Ftwo=  S(TU-C^2) +\coeff43 C^3 - \coeff12 U V_Y^2
 - C^2U +  \coeff{7}{24} U^3\, .
\eqn\aaahetone
$$
We see that with respect to $S$ the gauge field $C$ couples perturbatively
while with respect to $T$ it couples non-perturbatively in accord
with heterotic-heterotic duality.
Furthermore, $c_2(J_V)=-12$ and $V_Y$ couples neither 
to the dilaton nor to $C$. This suggests that $V_Y$ is the dual of a vector-tensor
which has  no Chern-Simons  coupling with $C$.
This is consistent with eq.~\Isum\
and the fact that the tensor and the $SU(2)$ originate from different
$E_8$ factors. Furthermore, since  $b=12$ this is consistent with \Presult.
However this vacuum can alternatively be viewed as 
$\IF_1+SU(2)_1+\tensor_1$
in the notation of the table. Eq.~\Isum\ then suggests the presence of a 
Chern-Simons coupling to the tensor field. Indeed, substituting
$$
t_1=S-\coeff12 U-\coeff12 T,\quad
t_2=T-V_Y-\coeff12 U,\quad
t_3=V_Y-\coeff12 U,\quad
t_4=C,\quad
t_5=U-3C,
\eqn\aaasustwo
$$
into \aaa\ we get
$$
\Ftwo=  S(TU-C^2) +\coeff43 C^3 - \coeff12 U V_Y^2
 - C^2U -\coeff12 C^2 T + C^2 V_Y +  \coeff{7}{24} U^3\, .
\eqn\aaahettwo
$$
This exhibits the Chern-Simons coupling $V_Y C^2$. The coefficients of 
$C^2 T$ and $C^2 U$ are no longer consistent with \Presult\ and $b=12$.
This descrepancy arises as \Presult\ has been derived under the assumption 
that the gauge field only couples to the fields in the perturbative
spectrum. It would be interesting to derive the coefficients without using
heterotic/Type II duality.

This feature can also be seen in our final example, vacuum 15
which could also be  viewed as $\IF_1+SU(2)_1+\tensor_1$.\foot{The 
polyhedron admits four triangulations with the
specified base which lead to the same prepotential. These
triangulations thus do not lead to distinct Calabi-Yau phases.}  
For one choice of heterotic variables the prepotential reads 
$$
\Ftwo=S(UT-C^2) + \coeff13 C^3 - \coeff12 U V_Y^2
+C^2 V_Y - \coeff12 U C^2 +\coeff7{24} U^3 \ ,
\eqn\bbbhetone
$$
whereas for a different choice we find 
$$
\Ftwo=S(UT-C^2) +\coeff13 C^3  - \coeff12 U V_Y^2
-\coeff12 UC^2 +\coeff12 T C^2 +  \coeff7{24} U^3 \ .
\eqn\bbbhetthree
$$
As in the previous example, the two different choices of heterotic coordinates
for the same type II vacuum correspond to the ambiguity of assigning the 
Chern-Simons couplings in eq. \Isum.  
\chapter{Discussion}

In this paper we   studied $d=4$ heterotic vacua compactified on
$K3\times T^2$ and their type II  duals. The latter are compactified
on the same elliptic Calabi-Yau  threefolds that are used in
F-theory to describe the non-perturbative behavior of six dimensional heterotic
vacua.
By computing the intersection numbers of $(1,1)$-forms of
Calabi-Yau  manifolds with $\IF_0$, $\IF_1$, $\IF_2$ and their toric blow-ups 
as bases we determined the couplings of the vector multiplets
with up to  $N_V=5$ in the associated prepotentials.
The consequences of
the (non-perturbative) properties of the heterotic string in $d=6$ were
displayed.

Using the techniques employed in the present work one should be able to
perform similar computations for other
$K3\times T^2$  heterotic vacua.  In particular 
shrinking an instanton in an $E_8$ with 
$n\leq 9$, leaves a terminal gauge group, which for $n=9$ is $SU(3)$ \DMW\
while for $n=8$ it is $SO(8)$ \CF. This is why in vacua 12, 22 and 25 we 
get two and in vacuum 26 four additional heterotic vector-multiplets. 
Furthermore, the massless matter of all vacua, which is determined
by the index theorem, should be reflected in the world-sheet 
instanton numbers, as explained in \BKKMI. Preliminary analysis
of the models considered here indicates that this is indeed the case.
A more detailed analysis might be worthwhile.

The results of this paper show that the type II prepotentials reproduce the
known perturbative couplings of the dual heterotic vacua.
This confirms the expectation that 
the four-dimensional heterotic--type II duality uses the
same Calabi-Yau manifold as the six-dimensional
heterotic--F-theory duality.
For vacua with  additional vector-tensor multiplets it would be
interesting to reproduce  the type II `predictions'  by an independent
heterotic computation. In particular a better understanding
of the heterotic interpretation of the flopped
Calabi--Yau phases is desirable.

The two possibilties of choosing heterotic variables
in vacua with gauge fields and tensor multiplets
motivated by the different factorizations
of the anomaly polynomial
appears to have an interesting interpretation
in terms of the `travelling' of a five-brane from one fixed
point to the other in  M-theory compactified on  $K3\times S^1/\ZZ_2$.
The form of the coupling after the 
detachment of  a five brane is under current investigation.

\appendix{A}{More flopped Calabi--Yau phases}

For completeness we list the prepotentials of some of the flopped Calabi--Yau
phases in this appendix.  In section~3.3 we discussed in detail vacuum 6
whose base is a blown up $\IF_0$ (or $\IF_1$) surface (fig.~3).
In eqs.~\Fzeroflop--\Fzeroflopf\ we also discussed the  flopped phase
corresponding to the vacuum build from the base on the right hand side
in fig.~3. The other possible flopped phase on the left
has
$c_2(J_\alpha)=\{92,36,24,82\}$ and
$$
\eqalign{
\Ftwo =\ \coeff76\,t_{{4}}^{3} + \coeff43\,t_{{1}}^{3}
+&\coeff12 t_{{2}}^{2}t_{{1}} + \coeff32\,t_{{2}}t_{{1}}^{2}
+ t_{{3}}t_{{1}}^{2}+ \coeff12 t_{{2}}^{2}t_{{4}} + 4\,t_{{1}}^{2}t_{{4}}
+\coeff32\,t_{{2}}t_{{4}}^{2} + t_{{3}}t_{{4}}^{2}+4\,t_{{1}}t_{{4}}^{2}\cr
+&\ \ t_{{2}}t_{{3}}t_{{1}}+t_{{2}}t_{{3}}t_{{4}}+3\,t_{{1}}t_{{2}}t_{{4}}
+2\,t_{{1}}t_{{3}}t_{{4}}\ .
}
\eqn\Fzeroflopp
$$
By setting $t_4=0$ one gets model 3 with prepotential \Foneone.
This corresponds to shrinking a four-cycle as indicated in fig.~3.
Going to heterotic variables via
$$
t_1=U-\BX,  \qquad
t_2=T-U, \qquad
t_3=S-\coeff12 U-\coeff12 T, \qquad
t_4=\BX\ ,
\eqn\Fzerofloppsubs
$$
we find
$$
\Ftwo=SUT + \coeff13 U^3  -\coeff16 \BX^3\ .
\eqn\Fzerofloppf
$$
We observe the same decoupling of $\BX$ as in \Fzeroflopf.

Instead of shrinking the four-cycle (or setting $t_4$ = 0) one can perform
a second flop on vacuum 6. This choice is indicated in the third row of fig.~3
and the `double flopped' phase is not a $K3$ fibration.
In deriving the data of the   phase
with the methods used above we encountered a subtlety.
The Mori cone and thus the K\"ahler cone is not simplicial.
Among the generators $l_i$ of the Mori cone we have a relation
$l_1+l_2=l_3+l_4$.
Introducing e.g.~the new generator $l_2-l_3$ we get a simplicial
cone. This cone has been used to derive the results;  we
also verified (as we have for all the models considered here) that the
instanton numbers are integers.\foot{ Here we do not worry about the 
possibility that the Mori cones of the Calabi-Yau hypersurface and the 
toric variety might differ
as when going to the
heterotic variables we are allowing for integer linear combinations
of the parameters anyway.
For a discussion of that point, see refs.~\refs{\PF,\BKKMI}.}
The data are
$c_2(J_i)=\{92, 102, 36, 82\}$ indicating that we do not have a $K3$ fibration
and
$$
\eqalign{
\Ftwo=& \coeff76 t_4^3 + 4 t_4^2 t_1+4 t_4 t_1^2+ \coeff43 t_1^3
+\coeff 32 t_4^2 t_3 + 3 t_4 t_1 t_3+ \coeff32 t_1^2 t_3 + \coeff12 t_4 t_3^2\cr
&+\coeff12 t_1 t_3^2 + \coeff92 t_4^2 t_1 + 9 t_4 t_1 t_2 + \coeff92  t_1^2 t_2
+3 t_4 t_2 t_3 +3 t_1 t_2 t_3\cr
&+ \coeff12 t_3^2 t_2 + \coeff92 t_4 t_2^2 + \coeff92 t_1 t_2^2
+ \coeff32 t_3 t_2^2 + \coeff32 t_2^3 \ .
}
\eqn\Fzeroflopppp
$$
For $t_4\to 0$ we get the flopped $\IF_1$ model with prepotential \Fonetwo,
whereas for
$t_1,t_4\to 0$ we get again the $\IP_2$ model.
The heterotic variables are
$$
t_1=W_X - \BX,  \qquad
t_2=U-W_X, \qquad
t_3=T-\coeff32 U, \qquad
t_4=\BX\ ,
\eqn\Fzeroflopppsubs
$$
which after substituting into \Fzeroflopppp\ results in
$$
\Ftwo=\coeff38  U^3  +\coeff12 UT^2 -\coeff16 \BX^3 -\coeff16 W_X^3\ .
\eqn\Fzeroflopppf
$$
As expected there is no dilaton and
two fields -- $\BX$ and $W_X$ -- decouple from $T$ and $U$.
For $W_X=0$ one recovers \Fonetwof.

One can also perform a flop on model 17 to arrive at a model with
$c_2(J_\a)=\{36,24,24,82,72\}$ and classical prepotential
$$
\eqalign{
\Ftwo&=t_5^3+t_5^2 t_2+ \coeff32 t_5^2 t_1+t_5 t_2 t_1+\coeff12 t_5 t_1^2
+t_5^2 t_3+t_5 t_2 t_3\cr
&+t_5 t_1 t_3+ \coeff72 t_5^2 t_4+2 t_5 t_2 t_4+3 t_5 t_1 t_4
+t_2 t_1 t_4+\coeff12 t_1^2 t_4+2 t_5 t_3 t_4\cr
&+t_2 t_3 t_4+t_1 t_3 t_4+ \coeff72 t_5 t_4^2 + t_2 t_4^2
+ \coeff32 t_1 t_4^2+t_3 t_4^2+ \coeff76t_4^3\,.}
\eqn\doubleupflop
$$
Setting $t_5=0$ we get \Fzerone.
Substituting
$$
t_1=T-\coeff12 U-V_Y,\quad
t_2=S-\coeff12 T-\coeff12 U,\quad
t_3=V_Y-\coeff12 U,\quad
t_4=U-\BX,\quad
t_5=\BX,
\eqn\doubleupflopsubs
$$
produces
$$
\Ftwo=STU - \coeff16 \BX^3 - \coeff12 U V_X^2+ \coeff7{24} U^3
\eqn\doubleupflophet
$$
(cf. \Fzeronef).

\ack

We would like to thank V.~Kaplunovsky for collaboration at an
early stage of this work and O. Aharony,
P.~Berglund, B.~de Wit, S.~Katz and A.~Klemm
for numerous fruitful discussions.
S.T.~and J.L.~thank the Physics Department in Heidelberg for their
hospitality.
S.T.~would also like to thank
the Physics Department of Tel Aviv University
and the Erwin Schr\"odinger Institute in Vienna for hospitality.

This  work is supported by GIF - the German-Israeli Foundation for
Scientific Research.
J.L.~is supported by a Heisenberg fellowship of the D.F.G.. 
J.L.~and S.T.~are supported by the European Community Research Programme 
under contract SCI-CT92-0789.
The work of J.S.~and S.Y.~was supported in part by the
Israel Academy of Science.

\refout
\end